\documentclass[sigconf,10pt]{acmart}
\usepackage[utf8]{inputenc}

\usepackage{algorithm}
\usepackage{float}
\usepackage{booktabs} 
\usepackage{algpseudocode}
\usepackage{calrsfs}
\usepackage[symbol]{footmisc}

\usepackage{lipsum}
\usepackage{caption}
\usepackage{subcaption}

\title{Network-Side Digital Contact Tracing on a Large University Campus}

\author{Matthew L. Malloy, Lance Hartung, Steve Wangen, Suman Banerjee}
\affiliation{%
   \institution{University of Wisconsin, Madison, United States}
}

\date{September 2021}

\AtBeginDocument{%
  \providecommand\BibTeX{{%
    \normalfont B\kern-0.5em{\scshape i\kern-0.25em b}\kern-0.8em\TeX}}}

\setcopyright{acmcopyright}
\copyrightyear{2022}
\acmYear{2022}
\acmDOI{10.1145/1122445.1122456}

\acmConference[MobiCom '22]{}{September 2022}{Planet Earth}
\acmBooktitle{Proceedings of The 28th Annual International Conference on Mobile Computing and Networking (MobiCom '22)}

\begin{abstract}
 We describe a study conducted at a large public university campus in the United States which shows the efficacy of network log information for digital contact tracing and prediction of  COVID-19 cases.  Over the period of January 18, 2021 to May 7, 2021, more than 216 million client-access-point associations were logged across more than 11,000 wireless access points (APs). The association information was used to find potential contacts for approximately 30,000 individuals. Contacts are determined using an AP colocation algorithm, which supposes contact when two individuals connect to the same WiFi AP at approximately the same time. The approach was validated with a truth set of 350 positive COVID-19 cases inferred from the network log data by observing associations with APs in isolation residence halls reserved for individuals with a confirmed (clinical) positive COVID-19 test result.  The network log data and AP-colocation have a predictive value of greater than 10\%; more precisely, the contacts of an individual with a confirmed positive COVID-19 test have greater than a 10\% chance of testing positive in the following 7 days (compared with a 0.79\% chance when chosen at random, a relative risk ratio of 12.6).  Moreover, a cumulative \emph{exposure score} is computed to account for exposure to multiple individuals that test positive.  Over the duration of the study, the cumulative exposure score predicts positive cases with a true positive rate of 16.5\% and missed detection rate of 79\% at a specified operating point.
 \end{abstract}

\begin{document}

\maketitle

\section{Introduction}
Digital contact tracing -- the use of digital devices such as mobile phones to establish potential epidemiological contacts -- has received significant attention as a potential tool in the COVID-19 pandemic.  Most proposed digital contact tracing approaches require installation of a mobile app or operating system level access which can be significant hurdles in countries where high public compliance is expected \cite{saw2021predicting, munzert2021tracking} and a fatal flaw elsewhere \cite{munzert2021tracking, time_nevada}. 

Conversely, network-side approaches to digital contact tracing do not require installation of a mobile application, operating system customization (i.e, Google and Apple's exposure notification system\cite{apple_google}), or any information collected on a client device.  Instead, they rely on network-side information such as connection logs to trace potential contacts.  As network-side approaches can be enabled by network operators without burden on end-users, they are attractive for large-scale, automated contact tracing. 

\begin{figure}[bt]
\includegraphics[width=0.4\textwidth]{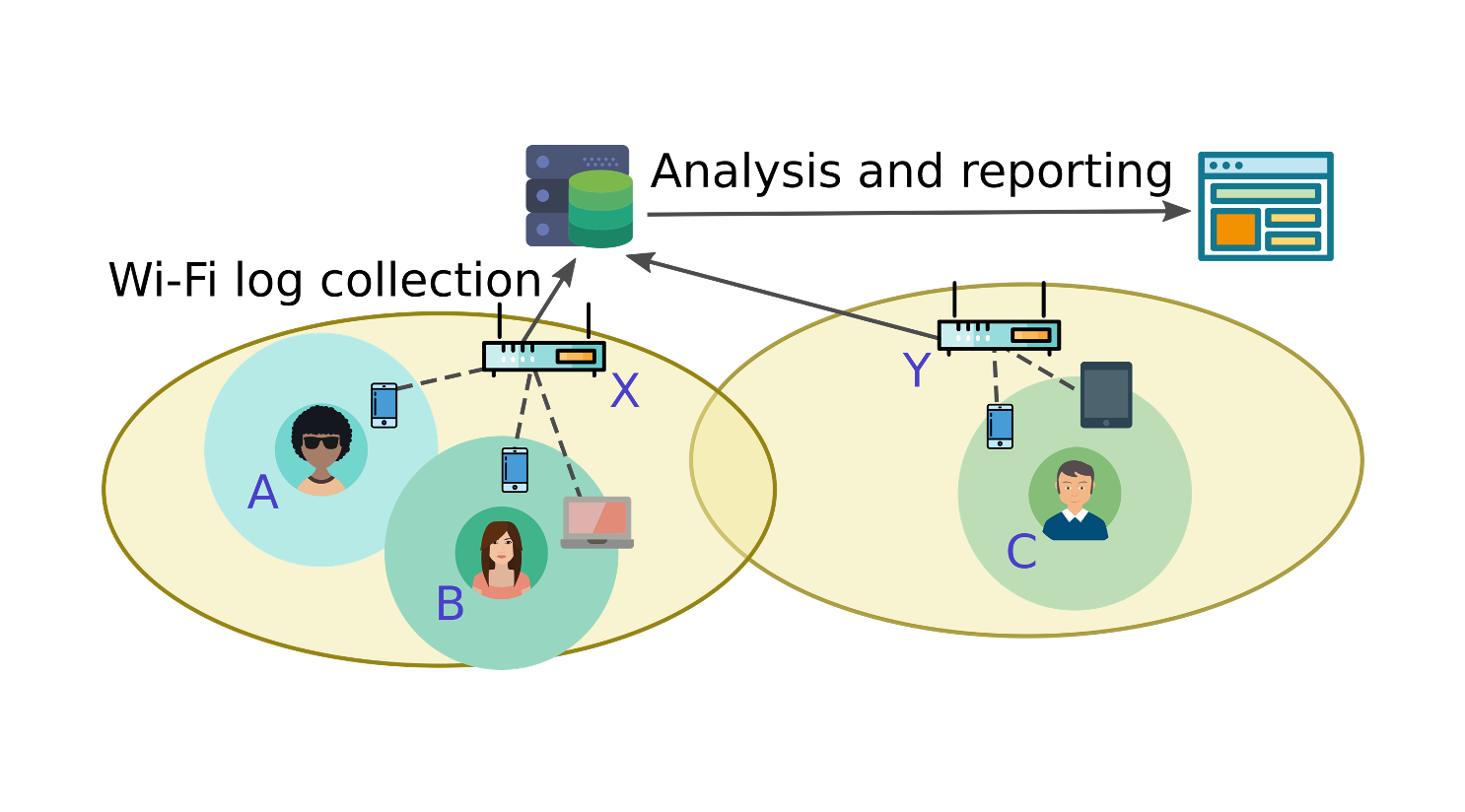}
\caption{Network-based digital contact tracing.  Two WiFi APs (denoted X and Y) are shown.  Two users (denoted A and B) connect simultaneously (i.e, colocate) to access point X, supposing epidemiological contact. }
\end{figure}

Access point (AP) colocation is the co-occurrence of users on the same WiFi access point (AP) at the same time.  As WiFi APs have limited range, AP-colocation is a proxy for physical colocation, which supposes epidemiological contact.  Like physical colocation, prolonged AP-colocation with an infected individual may correlate with increased risk of contracting the infectious disease.  To infer contacts and predict future infections, we present an \emph{AP-colocation algorithm}.  The algorithm generates a confidence score between two individuals that increases as the duration and number of AP-colocations increase.  More precisely, with an input corpus of connection logs, the algorithm outputs a time-varying weighted \emph{contact graph}.  The nodes in the graph correspond to individuals (or proxies for individuals, such as digital devices), and edges in the graph represent contacts between individuals. The weights of the edges are the confidence scores, which depend on the number of times two individuals colocate on a single AP and the number of other individuals connected to that AP.  The approach also assigns a cumulative \emph{exposure score} to individuals, which increases as multiple neighboring nodes in the contact graph test positive.

While approaches for digital contact tracing based on colocation have been proposed in the past \cite{malloy2020digital, trivedi2020empirical, trivedi2021wifitrace}, the primary contribution of this paper is a study of AP-colocation for digital contact tracing of COVID-19 at a large public university campus in the United States. 
The campus wireless network consists of more than 11,000 WiFi APs covering approximately one square mile and under normal circumstances serves around 50,000 students, employees, and visitors on a typical day.  The APs are located primarily inside residence halls, classroom buildings, libraries, dining halls, research and administrative buildings, shared outdoor spaces, and other spaces typical of a large university campus.  As students and employees move throughout the campus, their mobile devices connect and disconnect from the APs. The APs log connections and disconnections, creating a record of the approximate location of the user and others in their proximity.  
We apply the AP-colocation algorithm to a dataset of over 216 million WiFi association records collected over the duration of the Spring 2021 semester.  The resulting contact graph exhibits immense scale, and we report on its statistics.

To validate our approach, a truth set of positive (350) and negative (6,101) COVID-19 cases is inferred from the WiFi association dataset.  Positive cases are inferred by observing client-AP associations in dormitories reserved for COVID-19 isolation of individuals with a confirmed clinical positive.  Likewise, negative cases are inferred from associations in residence halls not reserved for isolation, which require a twice-a-week negative test result.  The ground truth dataset enables validation of the utility of the contact graph and the exposure scores for prediction of positive COVID-19 cases.  

Results indicate that the use of network log data and AP-colocation has a predictive value of greater than 10\% over the course of the study (above 16\% under some parameter choices resulting in limited scale).  More precisely, when tuned to return 2 contacts per positive case (on average), the returned contacts have greater than 10\% chance of having a confirmed positive COVID-19 result in the following 7 days.  This is contrasted with the 0.79\% chance of a positive result in the next 7-days when a contact is selected at random. To exploit when an individual is exposed to multiple positive cases, an exposure score is described and computed for each individual in the study.  For particular algorithm parameters, the cumulative exposure score predicts positive cases with a true positive rate of 16.5\% and missed detection rate of 79\%.

While the approach shows promise in settings such as a university or large corporate campus, there are significant shortcomings to using WiFi log data for digital contact tracing.  First, individuals must carry on their person a digital device that associates with the network.  Estimation of the percentage of individuals on the campus that do not associate with the enterprise WiFi was outside the scope of this study.  Second, it is possible and likely common for individuals that connect to the same AP to never come within a distance that supposes disease transmission.  As such, there are inevitable false positives (and missed detections), and the approach is best suited to establishing contacts associated with repeated and long term interaction between individuals.  As with any digital contact tracing, there are significant privacy considerations that must be addressed, and potential privacy risk must be contrasted with the benefit of such a system.  We discuss these trade-offs and note this study is meant to be a starting point for further conversation.  In light of these limitations, while the study suggests that digital contact tracing using network-side information can be effective, we recommend that it augment traditional contact tracing. 

Lastly, although this study was conducted at a large university campus, the ideas and techniques can be extrapolated to other settings in which network log data is collected.  In particular, both cellular network operators and entities in the digital advertising ecosystem collect the information required to implement network-side digital contact tracing in some form.  We refer the reader to \cite{malloy2020digital}. 

In summary, this paper proposes an approach for digital contact tracing based on network log data and describes a study conducted at a large public university campus in the United States.  To the best of our knowledge, this is the first study in which ground truth COVID-19 cases are used for validation of network-side contact tracing.

\section{Data}
\label{sec:data}
Data was collected at a large university campus in the United States during the Spring 2021 semester from January 18th 2021 through May 7th, 2021.  The data was collected from 11,964 physical Wi-Fi APs and 16 Aruba Networks enterprise network controllers. Automated log files containing association and disassociation event notifications were collected from the network controllers on a nightly basis. On average, a single day's log files contain 2,390,087 association events and 1,252,451 disassociation events during February 2021 (the first full month of the study in which students were present on campus and classes were in session).

Our study relies on the association records, which take the form \emph{(anonymized MAC address $i$, AP ID $k$, timestamp $t$)}, approximately localizing a user's device at a specific time.  Associations are only logged if the user also successfully authenticates with the network. This excludes records corresponding to randomized MAC addresses (i.e, those present in datasets collected from sniffing WiFi traffic such as probe requests outside an enterprise network). For security and auditing purposes, the network also records logs of authentication events. A sanitized dataset comprising 1-to-many mappings of securely hashed user IDs to MAC addresses was collected. Using the authentication dataset, we are able to group devices owned by the same user, which is crucial to interpreting our results. The anonymized mapping was used to convert the association records to the form \emph{(anonymized user ID $i$, AP ID $k$, timestamp $t$)}.

One limitation of the logging of the Aruba Network controllers is that approximately half of all recorded WiFi sessions ended without an explicit disassociation message, likely due to client devices roaming out of the AP's range. The absence of a clear session duration makes it challenging to implement contact tracing based on a precise calculation of the duration of AP colocation. Instead, in a method that also facilitates faster calculation, we discretize time into 15-minute epochs and record instances of \emph{(anonymized user ID $i$, AP ID $k$, epoch timestamp $t$)}. Repeated associations during an epoch are merged into a single record, and disassociation events are disregarded for this study.

Despite the scale and complexity, the data pipeline was remarkably reliable. Nonetheless, some level of attrition in collection was experienced, as software issues gradually prevented a subset of APs from reporting log messages. On three separate occasions, the APs from entire buildings were removed from our collection pool permanently as summarized in the table below. Affected buildings included dormitories, teaching spaces, and dining halls, but none of the isolation dorms were impacted. Visibility into 66\% of the campus was retained through the end of the Spring semester.

\begin{table}[h!]
    \begin{tabular}{|l|l|l|}
    \hline
    Date                & APs Reporting  & Buildings Included \\ \hline
    January 18, 2021    & 11,964 (100\%)  & 207 (100\%)        \\
    March 24, 2021      & 10,011 (84\%)   & 174 (84\%)         \\
    April 9, 2021       & 8,066 (67\%)   & 170 (82\%)         \\
    April 25, 2021      & 7,927 (66\%)   & 164 (79\%)         \\ \hline
\end{tabular}
\vspace{.1cm}
\caption{APs and buildings during the study duration.}
\end{table}

\subsection{Truth Set}
A truth set of positive and negative COVID-19 cases of residents of on-campus housing was inferred from the network traffic.  In the Spring 2021 semester, on-campus students were required to take twice-weekly rapid saliva-based COVID-19 tests.  Students who tested positive (and were residents of university housing) were required to move into one of five designated isolation dormitories.  Since all campus dormitories including the isolation dorms are covered by campus WiFi infrastructure, this allowed inference of positive cases based on extended and repeated observation of MAC address association to WiFi APs in the isolation dormitories.  

Likewise, an assumed negative was inferred by extended observation of a MAC address in campus residence halls not reserved for isolation.  Since students were tested twice weekly and positive cases were quickly moved to the isolation dormitories, an individual in a residence hall not reserved for isolation was an assumed negative.   Full details of the approach are included in the Appendix.  

Ultimately, the inferred ground truth data-set consisted of anonymized user IDs, an indicator if the user had an inferred positive test, and if so, the date and time at which they were observed to connect to an AP in the isolation dorm.

\section{Methodology}
\label{sec:method}

In this section we discuss the methodology used to predict potential exposure to infected individuals.  The approach requires first constructing a \emph{contact graph} followed by using the graph to predict contacts and ultimately new positive cases.

\subsection{Contact Graph}
To analyze and predict future cases from the WiFi association data, we construct a weighted, undirected, time-varying graph $G(t)$.  Following standard notation, a graph (or network) consists of a set of nodes $V$ and set of edges $E$.  An edge $e(t) \in E$ is a two element subset of the node set with an associated weight, $e(t) = (i,j,w) \in V \times V \times \mathbb{R}$.  


In an epidemiological contact graph, nodes correspond to an individuals (or surrogate for an individual, such as device identifier or a MAC address).  An edge represents a potential epidemiological contact between two individuals.  To control precision and recall, confidence scores -- denoted $w_{i,j}$ -- are assigned to edges. Larger weights represent a high potential for epidemiological exposure and higher likelihood for disease transmission. In the presentation of the algorithm we assume time has been discretized into epochs $t=1,2, \dots$.  The algorithm is described as follows.

\begin{algorithm}
\caption{AP Colocation Contact Graph  }\label{alg:graph}
\begin{algorithmic}[1]
\State{{\bf{parameters}}: look-back duration $\tau_g > 0$,  scaling parameter $\alpha \geq 0$ (default $\tau_g = 7$ days, $\alpha=1$)}
\State {\bf{input}}: AP associations: \emph{(device/user ID $i$, AP ID $k$, epoch $t' \leq t$)}
\State {$V= $ set of unique device/user IDs}
\State {\bf{for}} each AP $k$, each epoch $t' \in [t - \tau_g, t]$
\State \hspace{.4cm} $N_{k,t'} = $ number of IDs on AP $k$ on epoch $t' $
\State \hspace{.4cm} {\bf{for}} all pairs of IDs $(i,j)$ on AP $k$
\State \hspace{.8cm} $w_{i,j,k,t'} = \frac{1}{ N_{k,t'}^\alpha}$
\State $w_{i,j} = \sum_{k,t'} w_{i,j,k,t'}$ for all $(i,j)$
\State $E = \{(i,j,w_{i,j} ) \}$ 
\State \textbf{return} $G(t) = (V,E)$
\end{algorithmic}
\end{algorithm}
Algorithm \ref{alg:graph} takes client-AP associations as input, and computes a weighted (undirected) graph, similar to the approach proposed in  \cite{malloy2020digital}.  
The algorithm has two parameters: an optional scaling parameter $\alpha \geq 0$, and a look-back duration $\tau_g$.  Colocation between IDs prior to time $t-\tau_g$ are excluded from the contact graph $G(t)$.  $\tau_g$ is chosen to be longer than the incubation period for the disease, and excludes contacts that happened sufficiently far into the past.  The parameter $\alpha$ captures how the confidence score scales with number of devices connected to a common AP. If $\alpha = 1$, the score is inversely proportional with the number of devices on the AP, while when $\alpha = 0$, the edge weight between two users is the count of epochs and APs for which the users colocated.  A large value of $\alpha$ dilutes the effect of high volume APs, such as those found in dining halls.

Alg. \ref{alg:graph} assumes that time has been discretized into epochs.  The algorithm proceeds as follows: for each epoch $t'$ and access point $k$, the algorithm computes a corresponding weight between all IDs colocated on the AP.  Two users colocate on an AP if they both associate during a single epoch.  Weights are summed over valid epochs and APs to create edge weights.  The pairs of users that colocate on at least one epoch and their associated weights define the (time varying) graph $G(t)$.

\subsection{Predicting Positive Cases}
After construction of the contact graph, positive cases can be predicted.  If user $i$ has a (clinical) positive result, all neighbors of user $i$ (at the time of the positive result) with an edge weight above $\gamma$ are returned as predicted positives. 

When a disease is highly prevalent in a population, multiple contacts of an individual may test positive, increasing the chances of transmission to that individual.  In general, this is not captured by traditional contact tracing, as the contacts of each positive case are identified independently. 

To capture the potential predictive power of knowledge that multiple contacts have tested positive, we introduce the notion of an exposure score.  The exposure score is cumulative: for a single individual, the confidence scores associated with all positive contacts are summed.  More precisely, let $t_j$ be the time at which user $j$ tests positive.  The exposure score for user $i$ is defined as 
\begin{eqnarray*}
s_i(t) = \sum_{j \in N_p :t \in [t_j, t_j + \tau_s] }  w_{i,j}(t_j)
\end{eqnarray*}
where $N_p$ is the set of nodes that test positive during the study.  Since $w_{i,j} = 0$ for non-neighboring nodes, the sum is taken over neighbors of node $i$ that have a clinical positive test result.

\section{Results and Validation}

\subsection{AP-Colocation Graph}
The methodology of Sec. \ref{sec:method} was applied to a campus-wide dataset of more than 216 million client-AP associations over the duration of the study. As the graph is time-varying, we first report on the characteristics of the contact graph with $\tau_g = 7$ during the time period February 1, 2021 through February 8, 2021. We note that since a mapping between a MAC address and a user was only available for a subset of this data, the campus-wide contact graph was generated such that each node corresponds to a MAC address as opposed to an individual.

\begin{table}[h!]
\begin{tabular}{|l|l|}
\hline
{\bf Campus-wide contact graph } & count \\ \hline
client-AP associations & 16,124,734 \\ \hline
nodes, $|V(t)|$ & 47,415   \\ \hline
edge count, $|E(t)|$  & 2,242,934  \\ \hline
average degree ($w_{i,j} >0$) & 94.6 \\ \hline
\end{tabular}
\vspace{.1cm}
\caption{Statistics of the campus-wide contact graph $G(t)$, $\tau_g = 7$ days, $t = \mbox{February 8, 2021}$.  }
\end{table}

\subsection{Validation}
While the methodology was applied to a campus-wide dataset, only a subset of these associations correspond to individuals for which ground truth data was available.  The subset of data corresponding to these individuals was used to create a \emph{labeled} contact graph. The labeled contact graph $G$ consists of $6,451$ nodes (individuals), of which $350$ have a ground truth positive result during the course of the study.  The remaining $6,101$ individuals are assumed to be negative as described in Sec. \ref{sec:data} and the Appendix, for a positivity rate of 5.7\% over the duration of the study. 

For each individual that tests positive in the labeled dataset, the graph $G(t_i)$ was used to predict contacts, where $t_i$ denotes the time at which the individual tests positive.  A predicted positive contact is a neighbor of $i$ with a confidence score above a threshold $\gamma$; i.e, a predicted positive contact is a node $j$ such that $w_{i,j}(t_i) \geq \gamma$.  If such a neighbor tests positive in the following $\tau_p \in \{7,28\}$ days, a true positive (TP) event is recorded.  If such a neighbor does not test positive, a false positive (FP) event is recorded.

Let $P$ denote the set of individuals with a ground truth positive over the course of the study and $V(t)$ denote the set of individuals in the study at time $t$ (an individual is excluded from the study after testing positive).  The positive predictive value (PPV) is defined as
\begin{eqnarray} \label{eqn:ppv}
\mbox{PPV} &=& \frac{\mbox{TP}}{\mbox{TP} + \mbox{FP}} \\ \nonumber
&=&  \frac{ \sum_{i \in P} \sum_{j \in P } \mathbb{I}\{ w_{i,j}(t_i)  \geq \gamma \cap  t_i < t_j \leq t_i + \tau_p  \}}{\sum_{i \in P}  \sum_{j \in V(t_i)} \mathbb{I}\{ w_{i,j}(t_i)  \geq \gamma  \}}. \nonumber 
\end{eqnarray}
where TP and FP represent the count of false positives and true positives over the duration of the study, and $\mathbb{I}\{\cdot \}$ is the indicator function.

\begin{table}[h!]
\begin{tabular}{|l|l|}
\hline
 {\bf Validation Set} & count \\ \hline
individuals (nodes) at start, $|V(0)|$ &  $6,451$  \\ \hline
individuals (nodes) at end, $|V(t_{e})|$ & $6,101$  \\ \hline
positive cases (nodes), $|P |$ & $350$   \\ \hline
\end{tabular}
\caption{Statistics of the validation set.}
\end{table}

For comparison, the positive predictive value of contacts chosen at random was calculated.  Again let $i$ and $j$ index the individuals (nodes), and $t_i$, $t_j$ denote the respective time at which the individual tests positive, then  
\begin{eqnarray*}
\mbox{PPV}_{\mbox{rand}} = \frac{1}{|P|} \sum_{i \in P} \frac{  \sum_{j \in P} \mathbb{I}\{ t_i < t_j \leq t_i + \tau_p \}}{ |V(t_i)| }.
\end{eqnarray*}

\begin{table}[h!]
\begin{tabular}{|l|l|l|l|}
\hline
& parameters (days) & PPV & scale  \\ \hline
Alg. 1 & $\tau_p = 7, \tau_g = 7$ & 5.0\% & 5.2 \\ \hline
Alg. 1 & $\tau_p = 7, \tau_g = 7$ & 10.0\% & 1.7 \\ \hline
Alg. 1 & $\tau_p = 7, \tau_g = 7$ & 12.5\% & 0.9 \\ \hline
Alg. 1 & $\tau_p = 28, \tau_g = 7$ & 5.0\% & 8.4 \\ \hline
Alg. 1 & $\tau_p = 28, \tau_g = 7$ & 10.0\% & 2.1 \\ \hline
Alg. 1 & $\tau_p = 28, \tau_g = 7$ & 12.5\% & 1.1 \\ \hline
$\mbox{PPV}_{\mbox{rand}}  $ & $\tau_p = 7$ & $0.79\%$ & NA \\ \hline
$\mbox{PPV}_{\mbox{rand}}$  & $\tau_p = 28$ & $2.12\%$ & NA\\ \hline
\end{tabular}
\caption{ \label{tab:res} Validation results at various operating points, compared with predicted contacts chosen at random. }
\end{table}

\begin{figure}[htb]\centering
   \includegraphics[scale=0.45]{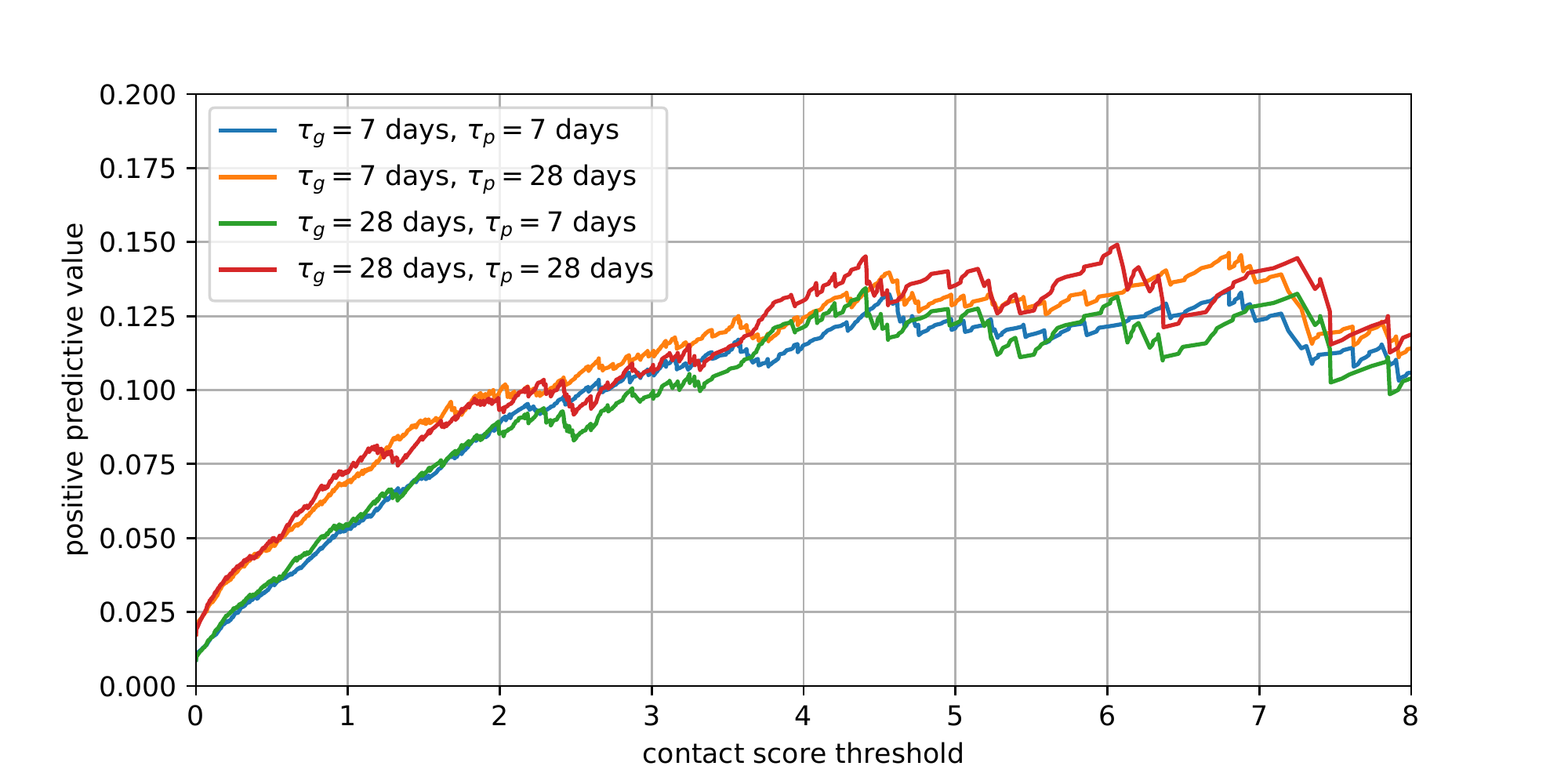} \\
   \caption{\label{fig:pp1}  Percent of predicted contacts that test positive in the $\tau_p$-days ($\tau_p \in \{7,28\}$) after a positive case as a function of the threshold $\gamma$, denoted PPV.  Contact graph with $\tau_g \in \{7,28\}$.  $\alpha = 1$.}
\end{figure}

\begin{figure}[htb]\centering
   \includegraphics[scale=0.45]{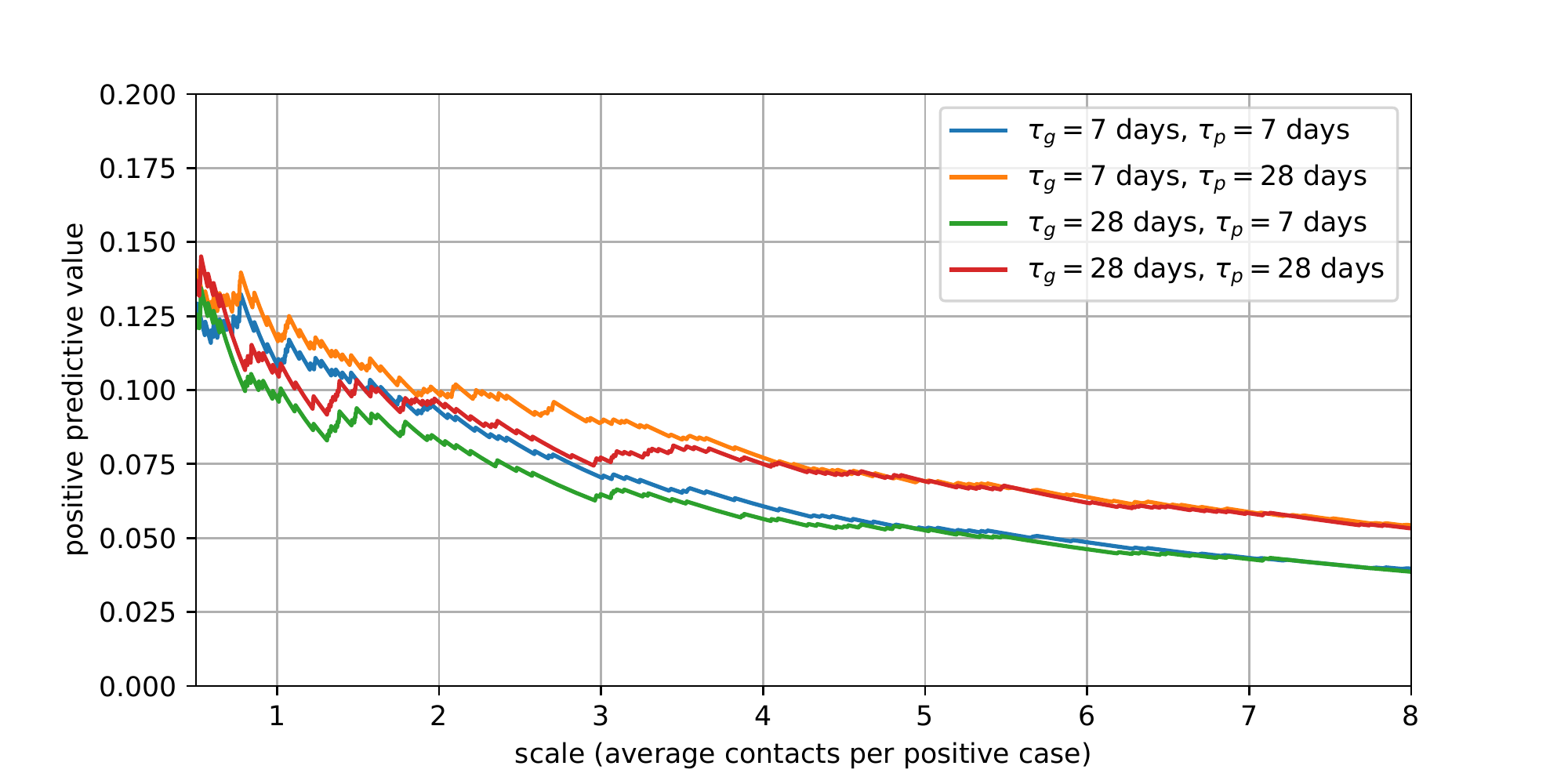} \\
  \caption{\label{fig:pp2} 
Scale (average number of predicted contacts) vs. positive predictive value for $\tau_p \in \{7,28\}$ and $\tau_g \in \{7,28\}$.  $\alpha = 1$. }
\end{figure}

Scale is defined as the average number of contacts that are returned for a given contact score threshold. More specifically, 
\begin{eqnarray*}
\mbox{scale} = \frac{\mbox{TP}+\mbox{FP}}{|P|} = 
 \frac{ \sum_{i \in P}  \sum_{j \in V(t_i) } \mathbb{I}\{ w_{i,j}(t_i)  \geq \gamma \}}{ |P|  }.
\end{eqnarray*}

\begin{figure*}[h]\centering
  \begin{tabular}{cc}
   \includegraphics[scale=0.41]{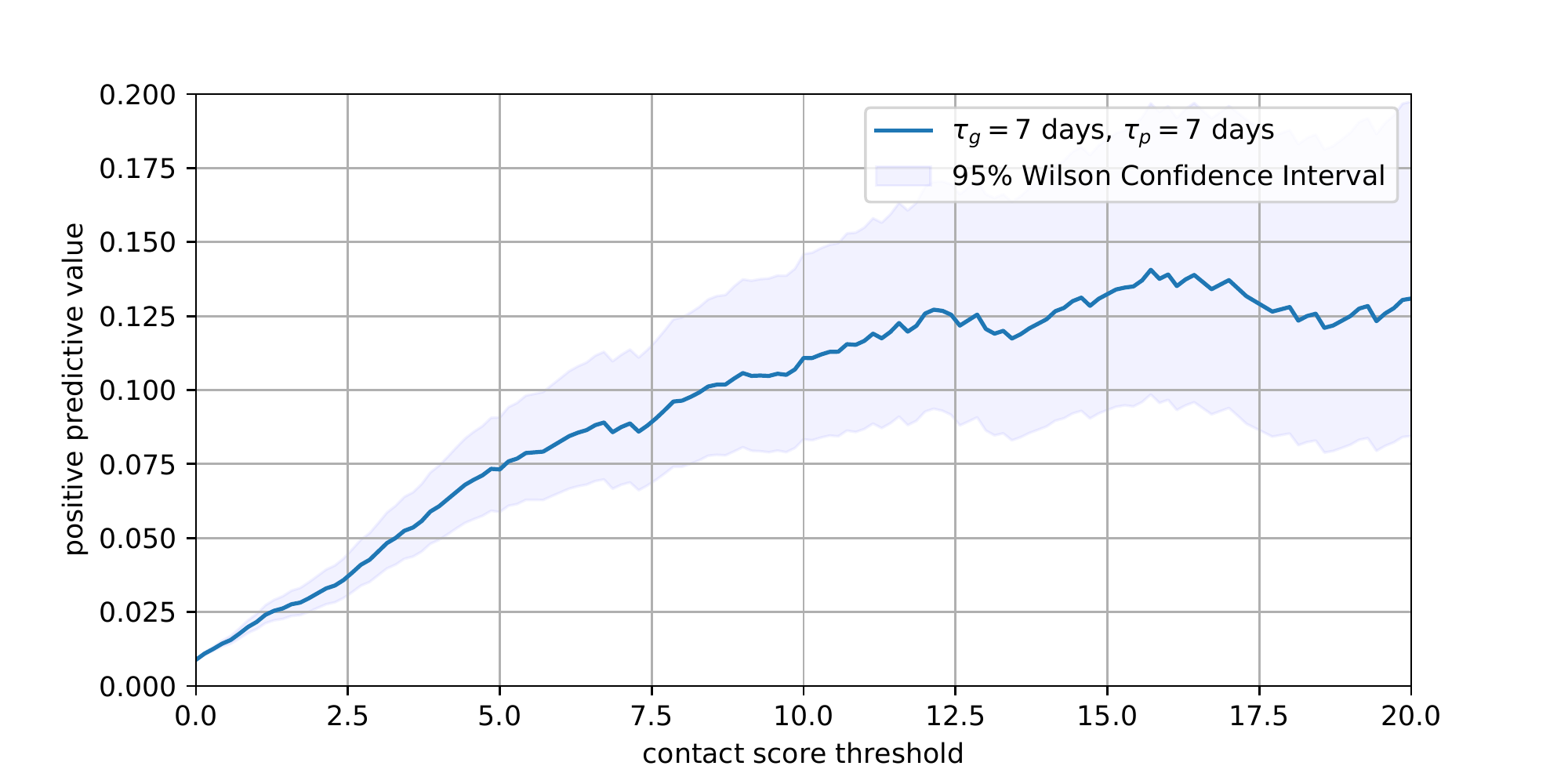} & \includegraphics[scale=0.41]{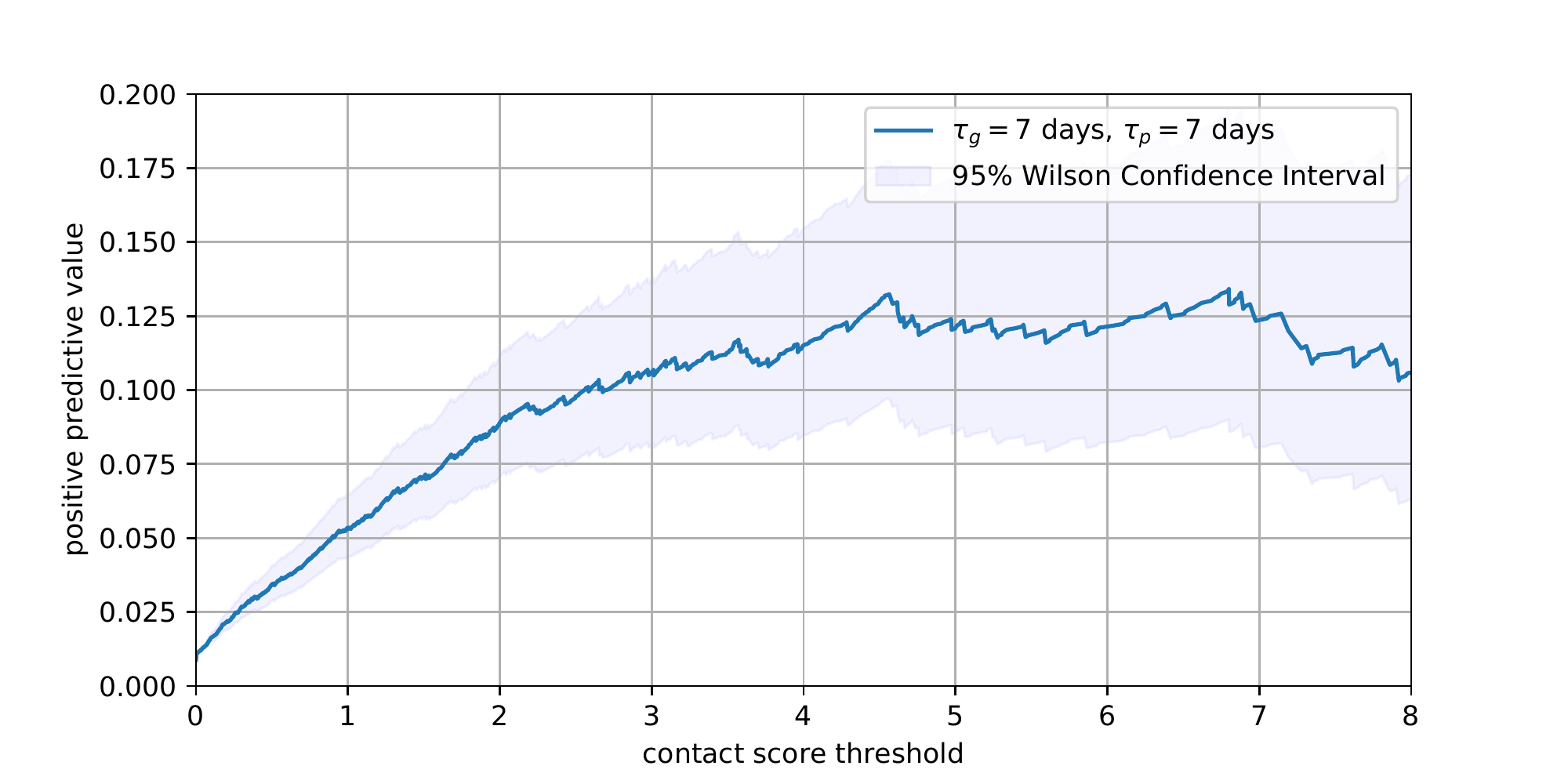} \\
   (a) & (b) \\
   \end{tabular}
  \caption{\label{fig:pp11} Percent of predicted contacts that test positive in the $\tau_p = 7$ days after a positive case, as a function of the threshold $\gamma$.  $\tau_g = 7$ days.  A $95$\%-Wilson confidence interval is shown. Figure a, $\alpha = 0$.   Figure b, $\alpha = 1$.}
\end{figure*}

\begin{figure*}[h]\centering
  \begin{tabular}{cc}
   \includegraphics[scale=0.41]{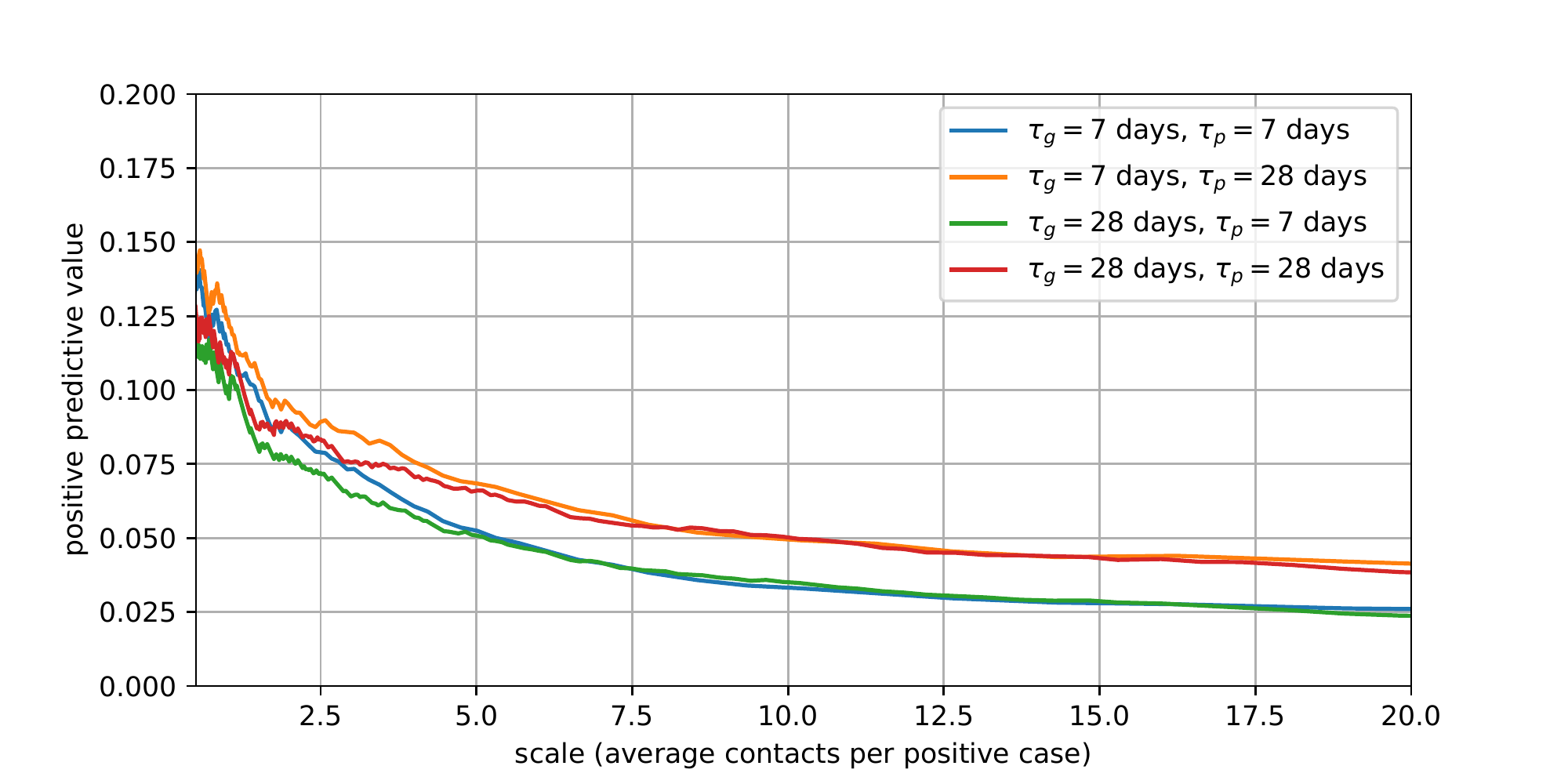} & \includegraphics[scale=0.41]{figs/scale_ppv_g1.pdf} \\
   (a) & (b) \\
   \end{tabular}
  \caption{\label{fig:pp22} 
Scale (average number of predicted contacts) vs. positive predictive value for $\tau_p \in \{7,28\}$ and $\tau_g \in \{7,28\}$. Figure a, $\alpha = 0$.   Figure b, $\alpha = 1$. Note that at some scale (i.e, 5), $\alpha =1$ has a higher positive predictive value than $\alpha = 0$.  }
\end{figure*}

Scale and PPV are shown in Fig. \ref{fig:pp1} and Fig. \ref{fig:pp2} for the following parameter settings: $\alpha = 1$, $\gamma \in [0, 20]$, $\tau_g \in \{7, 28\}$ days, $\tau_p \in \{7, 28\}$ days.  Additional results for a variety of parameter choices are shown in Fig. \ref{fig:pp11} through Fig. \ref{fig:pp22}.

\begin{figure*}[h]\centering
  \begin{tabular}{cc}
   \includegraphics[scale=0.41]{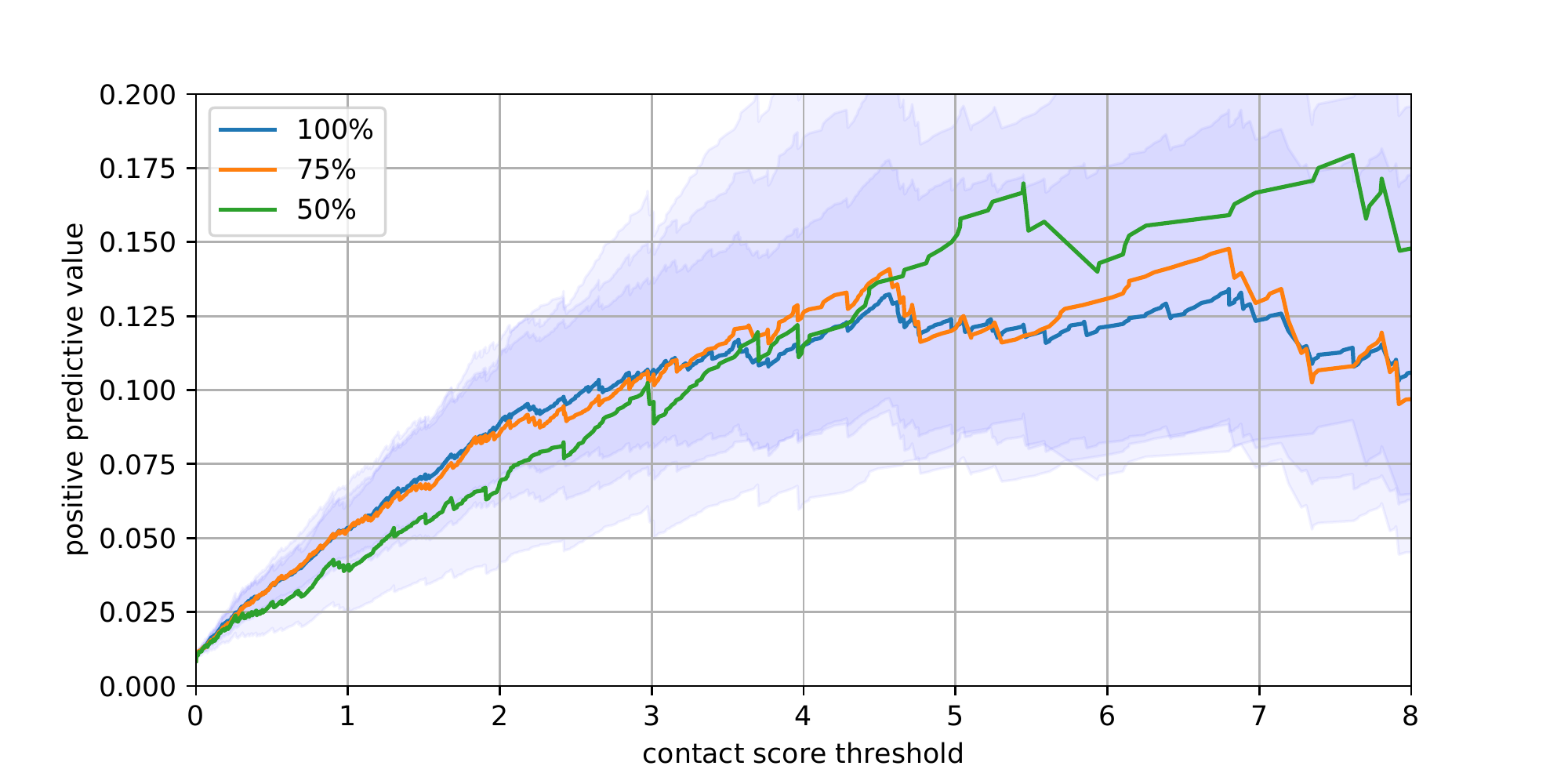} & \includegraphics[scale=0.41]{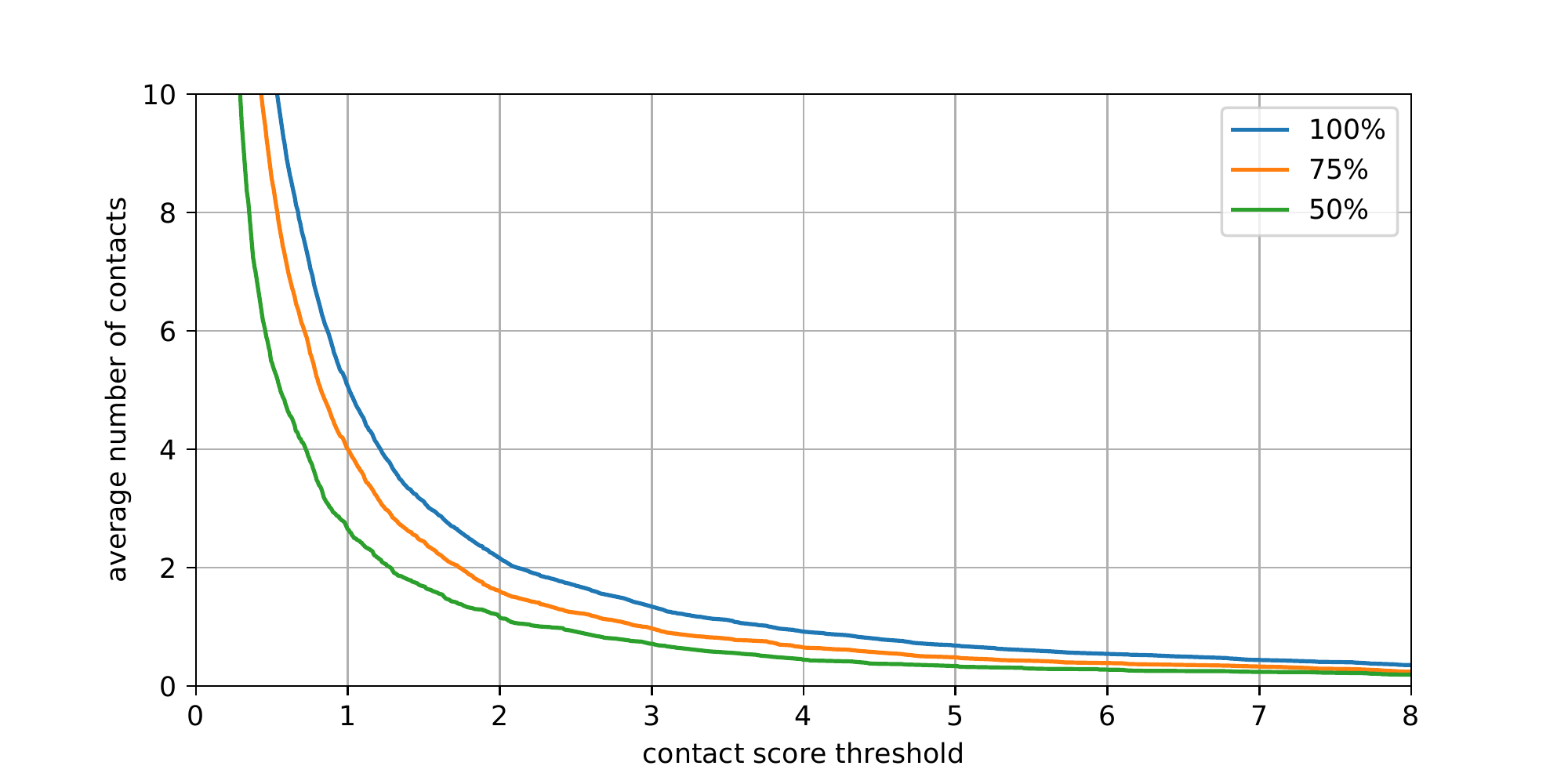} \\
   (a) & (b) \\
   \end{tabular}
  \caption{\label{fig:sensitivity} 
Sensitivity analysis when 100\%, 75\%, and 50\% of the original $6,451$ users participate in the study. 95\%-Wilson confidence intervals are shown.  Note the positive predicted value is largely unaffected by decreased participation, but the average number of contacts (per positive case) decreases accordingly.  $\tau_p = 7$ and $\tau_g =7 $.  $\alpha = 1$. }
\end{figure*}

Fig. \ref{fig:sensitivity} show analysis of the sensitivity to participation in the study.  In particular, of the $6,451$ individuals, 25\% and 50\% were excluded at random, and positive predictive value and scale analysis was repeated.  This resulted in a study size of $4,840$ users and $256$ positives for 75\% participation, and $3,226$ users and $159$ for 50\% participation (contrasted with $6,454$ and $350$ for full participation). 
The positive predictive value does not change significantly since both the numerator and denominator of (\ref{eqn:ppv}) are reduced by approximately the same amount.  The scale (average returned contacts per positive) decreases proportional to the number of devices in the study.

\begin{figure}[htb]\centering
   \includegraphics[scale=0.55]{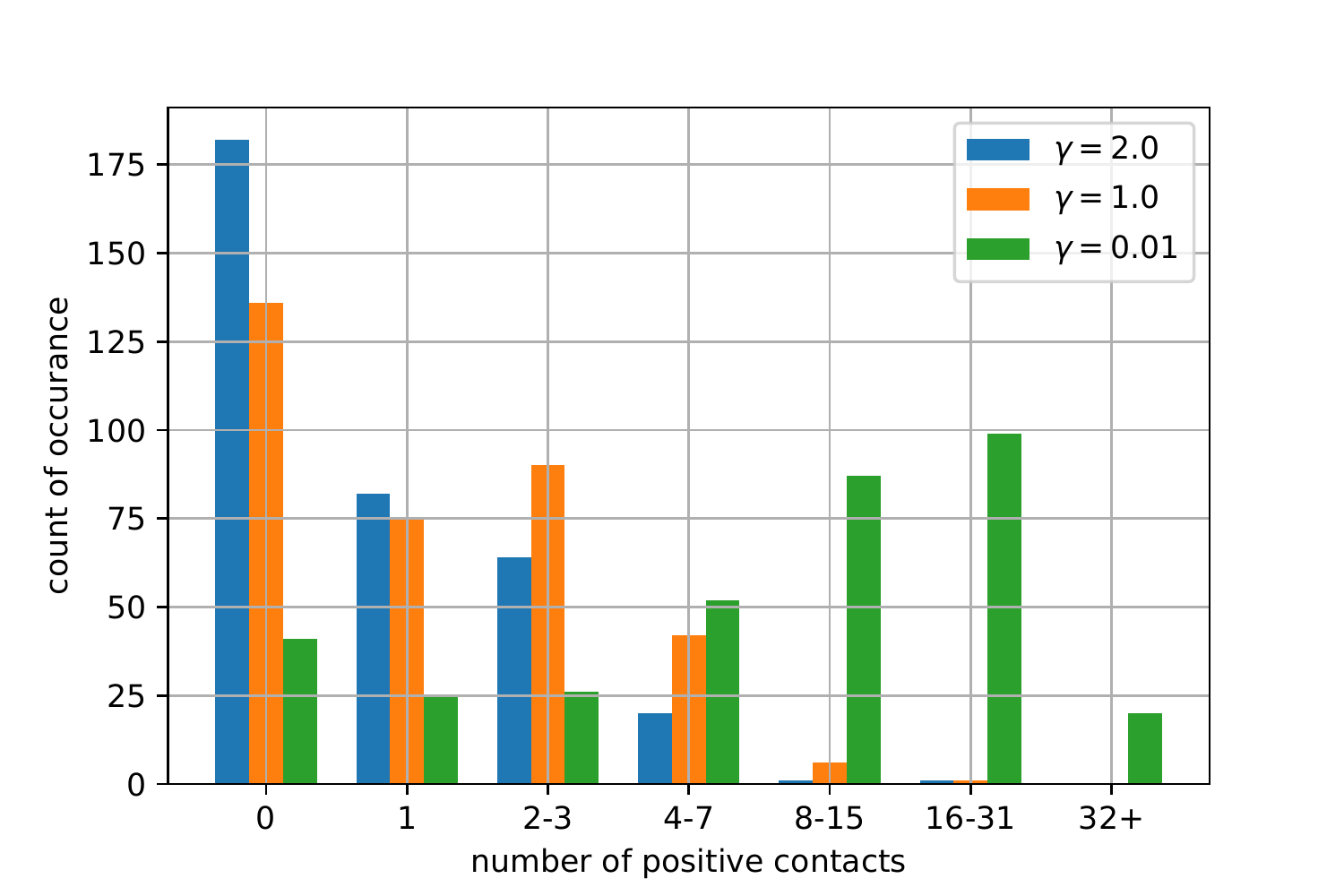} \\
  \caption{\label{fig:pos_degree}
Histogram showing the number of plausible transmissions from each positive case. The $x$-axis indicates number of contacts with an edge weight $w \geq \gamma$ and a confirmed positive in the following $\tau_p = 7 $ days, and the $y$-axis shows the count of occurrences out of the 350 positive cases.  $\tau_g = 7 $ and $\alpha = 1$. }
\end{figure}

Fig. \ref{fig:pos_degree} shows a histogram of the number of plausible transmissions, denoted $\widehat{R}$.  A \emph{plausible transmission} is the number of contacts of a positive case (with edge weight $w\geq \gamma$) that test positive in the following $\tau_p$ days.  More precisely, let $i$ index a user with a confirmed positive at time $t_i$.  Then the plausible transmissions for case $i$ is given as
\begin{eqnarray*}
\widehat{R}_i = \sum_{j \in P } \mathbb{I}\{ w_{i,j}(t_i)  \geq \gamma \cap  t_i < t_j \leq t_i + \tau_p  \}.
\end{eqnarray*}
Fig. \ref{fig:pos_degree} shows histogram of $\widehat{R}_i$ over the 350 positive cases. 

\begin{figure*}[h] \centering
\begin{tabular}{cc}
\includegraphics[scale=0.41]{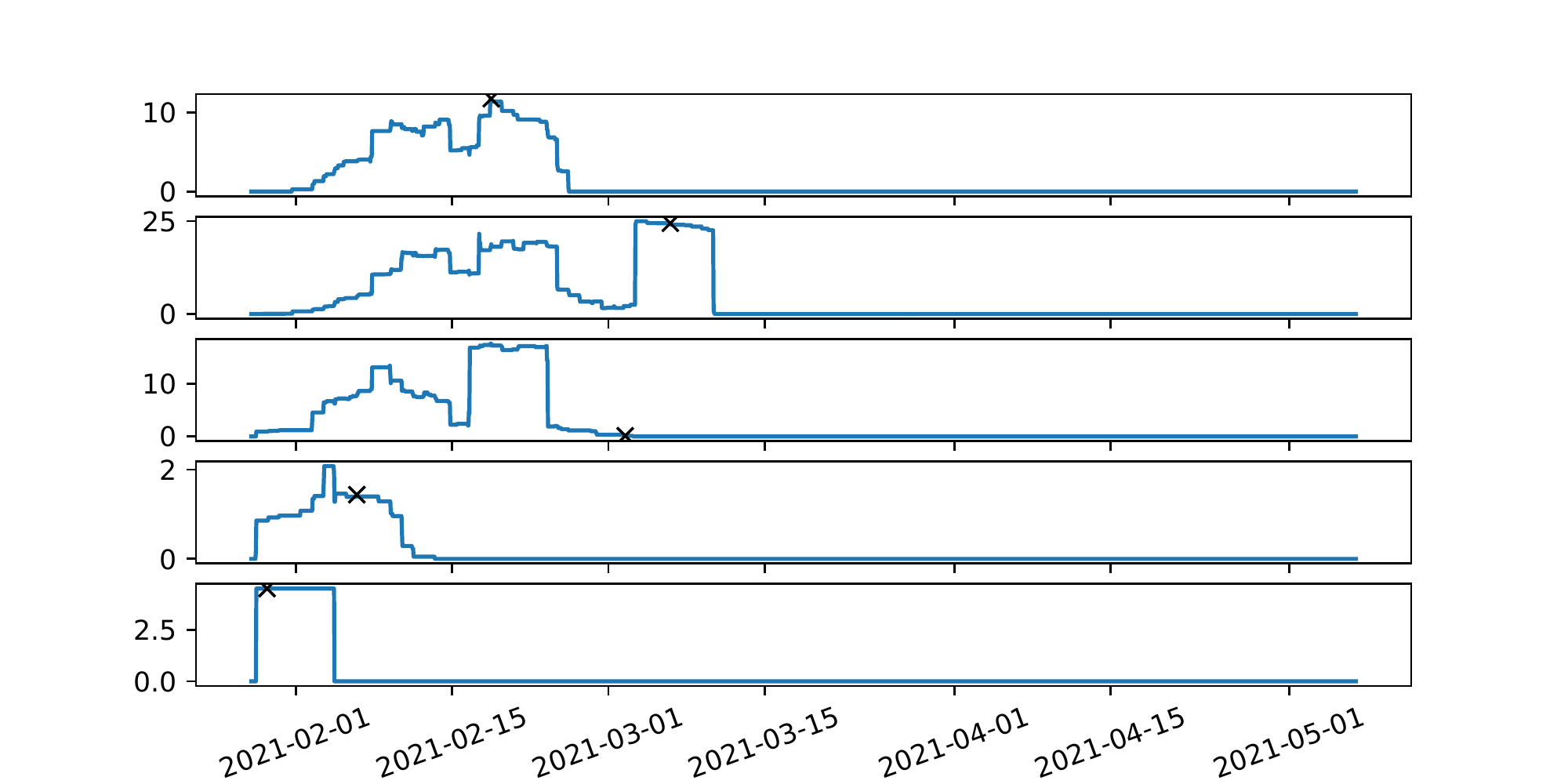} &
\includegraphics[scale=0.41]{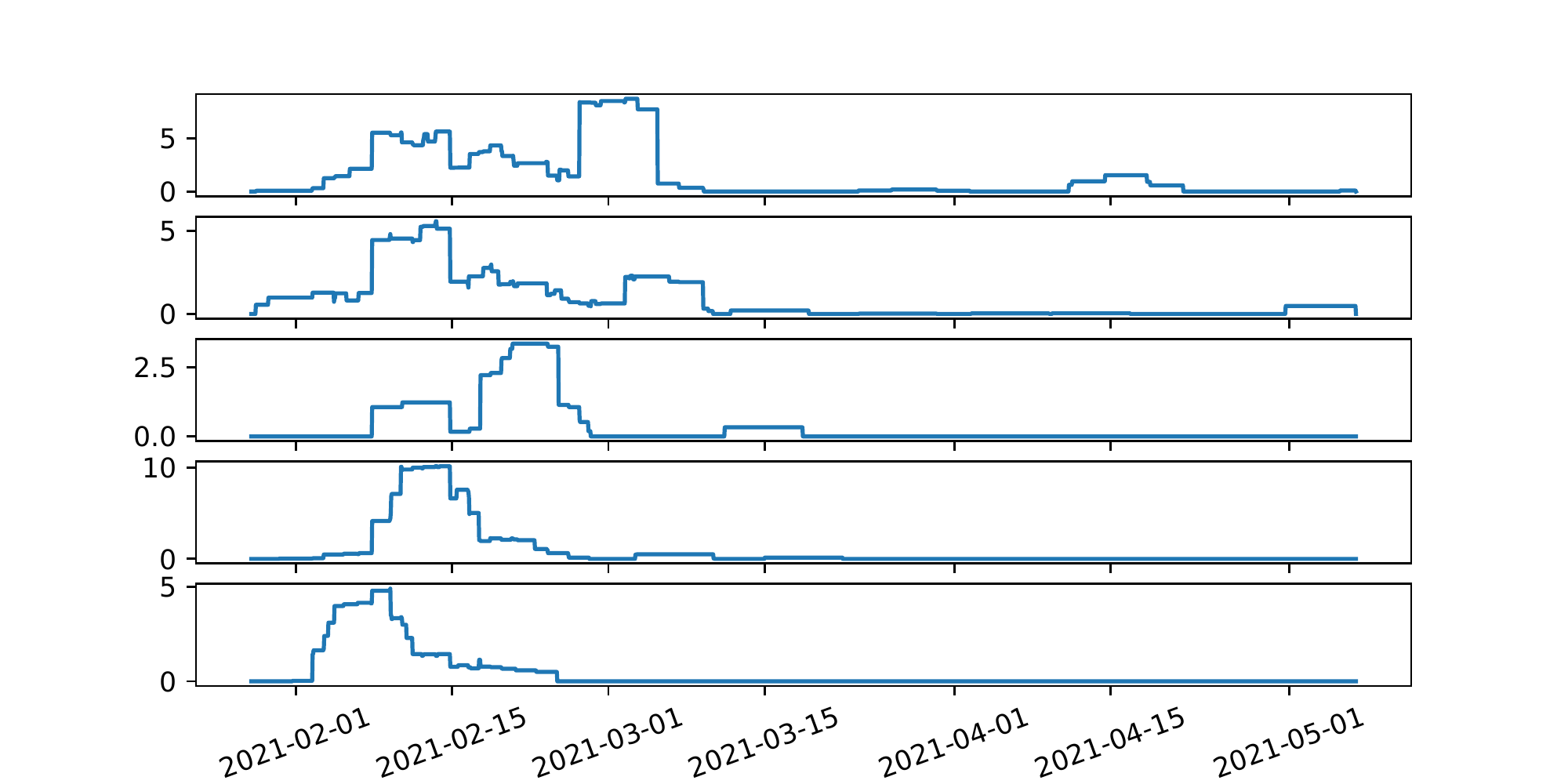} \\ 
(a) & (b) 
\end{tabular}
\caption{\label{fig:pos_examples} Exposure scores $s_i(t)$ of five users that test positive (Figure a) and negative (Figure b) during the course of the study.  The date of the positive test result is indicated by an `\texttt{x}'.  Note the scale of the y-axis is different for each sub-plot. $\tau_s = 7$ days.}
\end{figure*}

In addition to the validation of the confidence score produced by the graph, the exposure score as a predictor of a future positive COVID-19 test was validated.  The \emph{exposure score} aims to capture the potential predictive power of knowledge that multiple contacts of an individual have tested positive.  The exposure score $s_i(t)$ of ten individuals (chosen randomly) over the course of the study are shown in Fig. \ref{fig:pos_examples}.  The left figure in Fig. \ref{fig:pos_examples} shows the score corresponding to individuals that test positive.  The date of the positive test is shown on the plot.  In all cases, the exposure score is elevated prior to positive result.  The right figure shows example traces of individuals that do not test positive.  Note that the y-axis is scaled for each subplot.

The cumulative exposure score was also used to predict positive cases. For each individual, a positive prediction was declared at the first time the exposure score exceeded a threshold $\gamma$.  If the time of the prediction was \emph{after} the individual had a ground truth positive, the individual was excluded.  If the time of the prediction preceded a ground truth positive test by less than $\tau_p$ days, a true positive (TP) was recorded; if the time of the prediction preceded a ground truth positive test by more than $\tau_p$ days, or the individual did not have a ground truth positive over the course of the study, a false positive (FP) was recorded.  Likewise, if the confidence score of the individual was below $\gamma$ for the duration of the study and the individual did not test positive, a true negative (TN) was recorded. If the confidence score of the individual was below the threshold for the duration of the study and the individual tested positive, a false negative (FN) was recorded.  The true positive rate is:
\begin{eqnarray*}
\mbox{true positive rate} = \frac{\mbox{TP}}{\mbox{TP} + \mbox{FP}}. 
\end{eqnarray*}
Likewise, the missed detection rate is the percentage of total positive cases that are not predicted
\begin{eqnarray*}
\mbox{missed detection rate} = \frac{\mbox{FN}}{\mbox{TP} + \mbox{FN}}.
\end{eqnarray*}

\begin{figure}[htb] \centering
\includegraphics[scale=0.45]{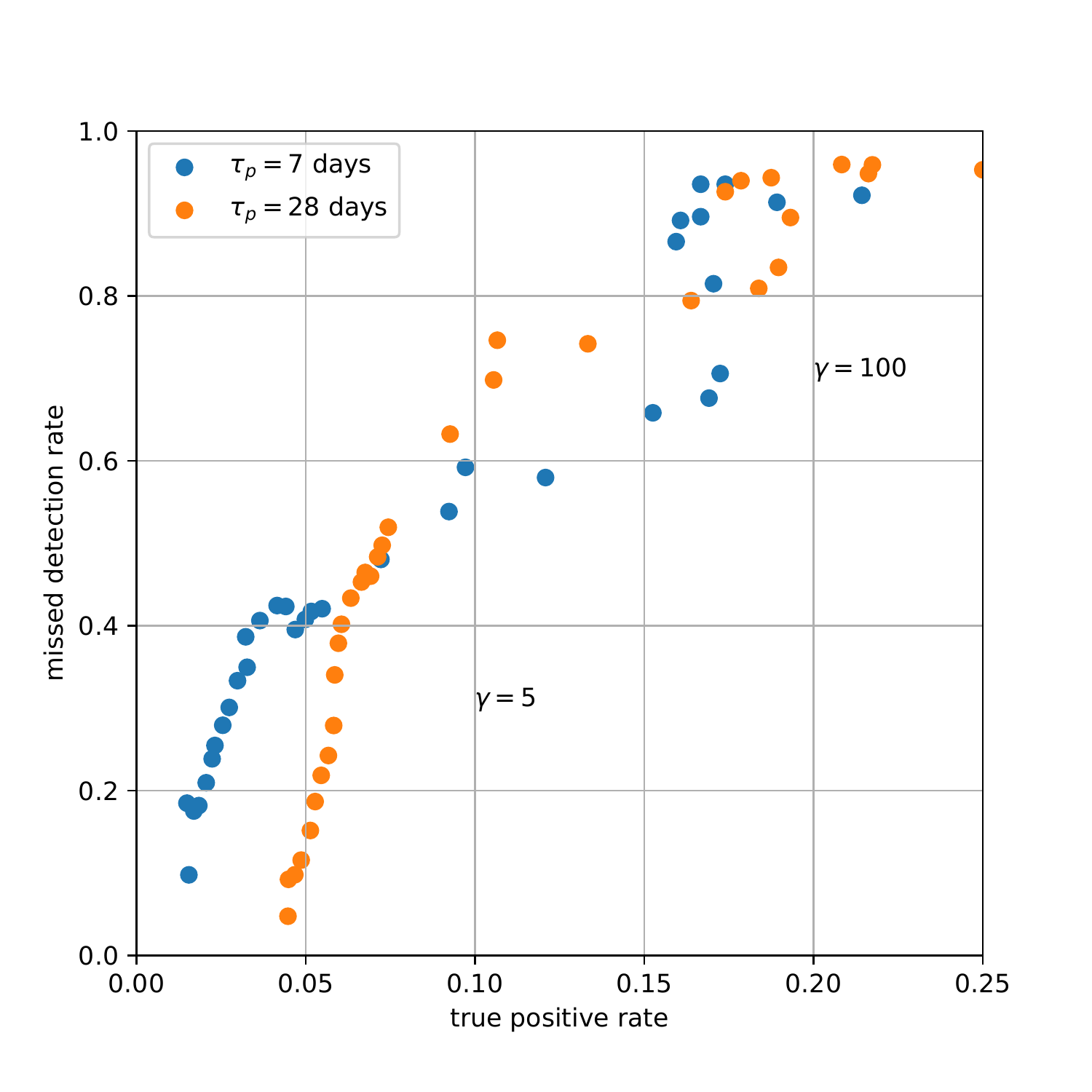} 
\caption{\label{fig:roc} Receiver operating characteristic curve for the exposure score $s_i(t)$ over the duration of the study.  The true positive rate plotted against missed the detection rate for values of $\gamma \in [1, 200]$, for $\tau_p = \{ 7, 28\} $ days.  $\tau_g = 7$, $\tau_s = 7$. }
\end{figure}

Fig. \ref{fig:roc} show the true positive rate and missed detection rate for a number of values of $\gamma$ over the duration of the study. Note that $\gamma$ can be used to select an appropriate operating point:
the approach can operate with a true positive rate of 16.5\% and missed detection rate of less than 80\%, or, for example, a missed detection rate below 20\% and true positive rate above 5\%. 

\begin{figure*}[h] \centering
\begin{tabular}{cc}
\includegraphics[scale=0.41]{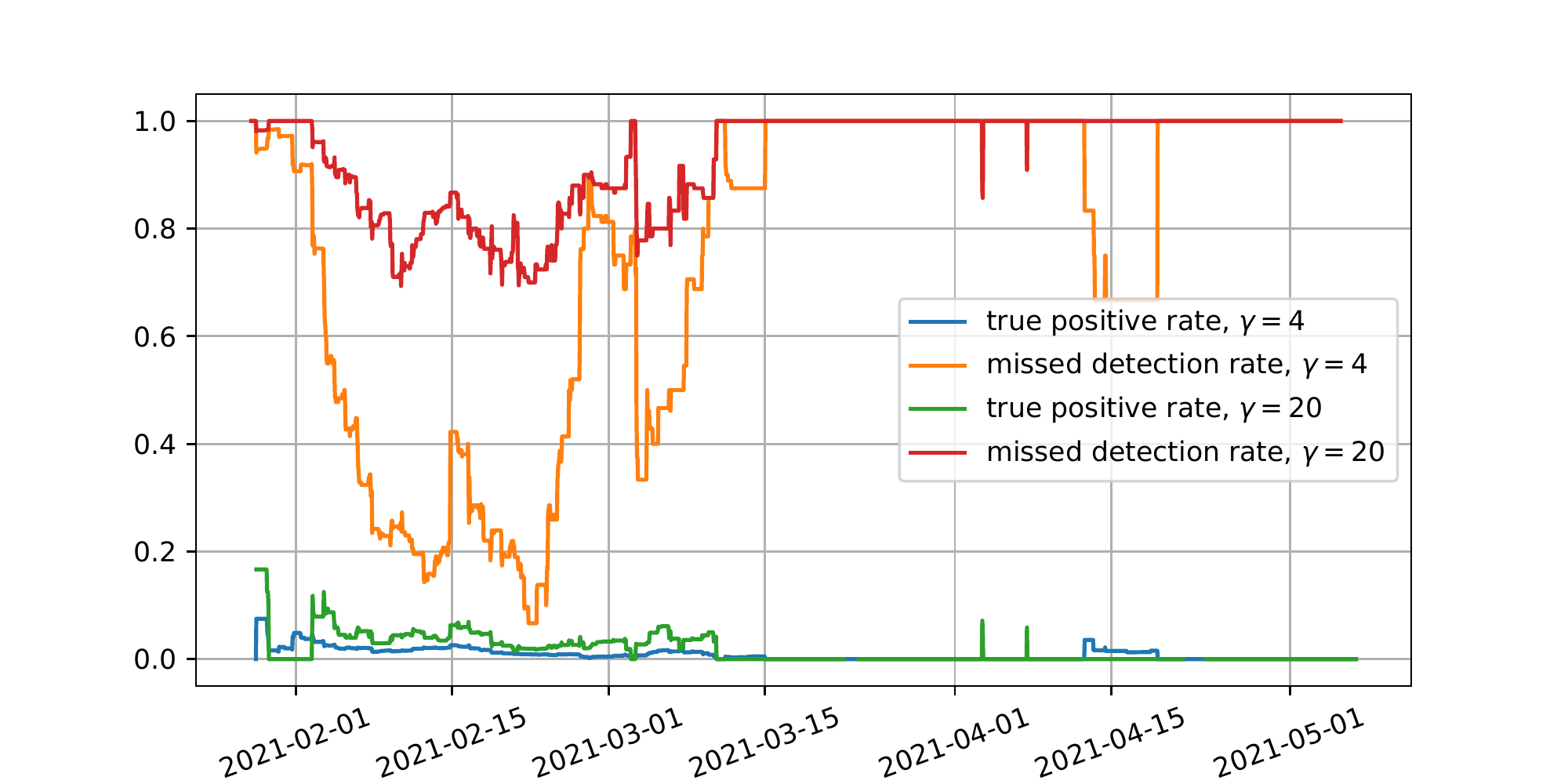} & 
\includegraphics[scale=0.41]{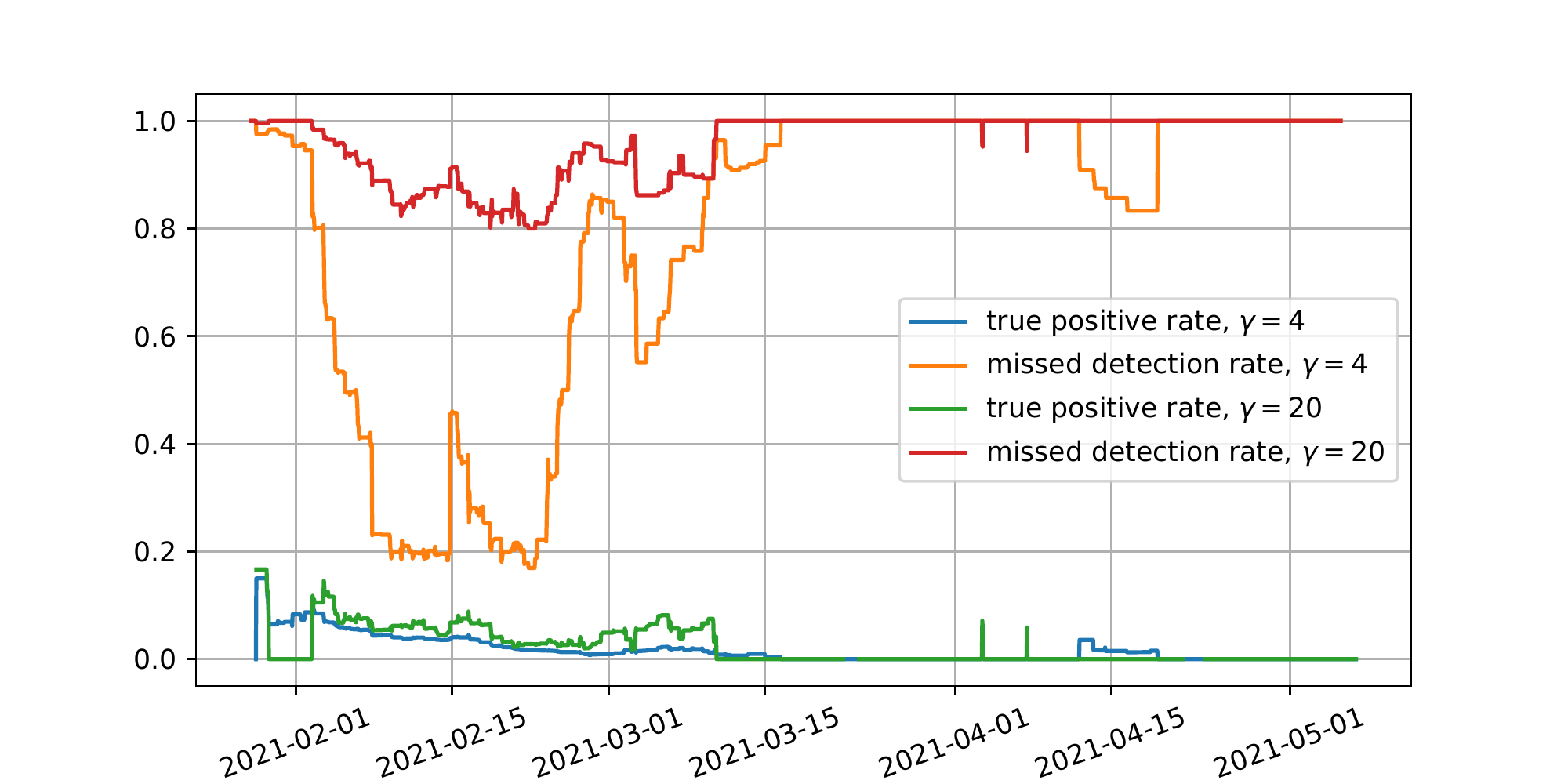} \\
(a) & (b) 
\end{tabular}
\caption{\label{fig:TP_rate} True positive rate -- predicted positive individuals that test positive in $\tau_p = 7$ days after date, divided by the total number of predicted positive individuals.  Missed detection rate -- predicted negatives that test positive in $\tau_p$ days after the date, divided by count of positive tests in $\tau_p$ days after the date.  A predicted positive is an individual with a cumulative exposure score above $\gamma \in \{4, 20\}$, and a predicted negative is a an individual with a cumulative exposure score below $\gamma \in \{4, 20\}$.  $\tau_g = 7$, $\tau_s = 7$. Figure a, $\tau_p = 7$. Figure b $\tau_p = 28$.}
\end{figure*}

\begin{figure*}[h] \centering
\begin{tabular}{cc}
\includegraphics[scale=0.41]{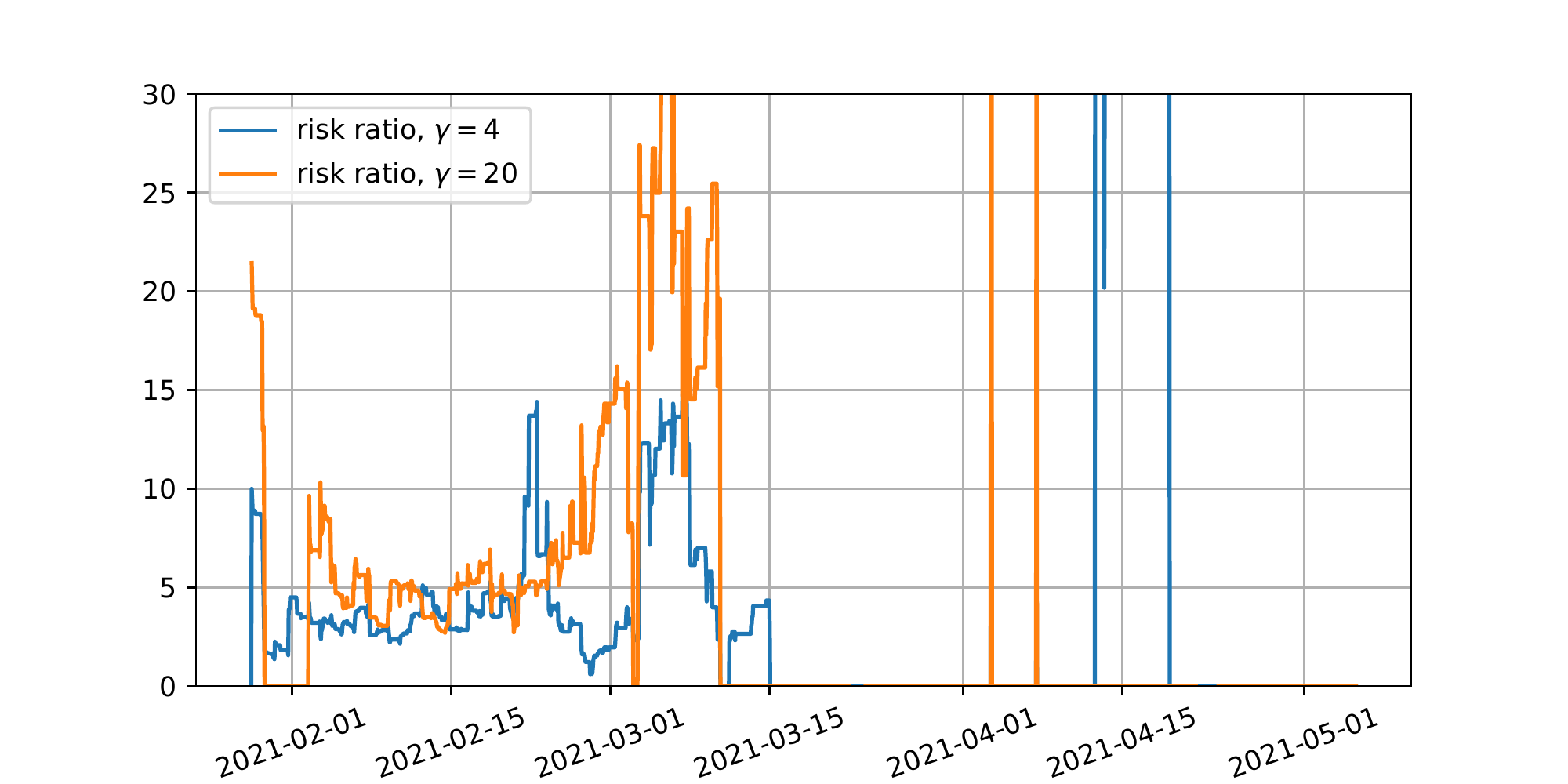}
\includegraphics[scale=0.41]{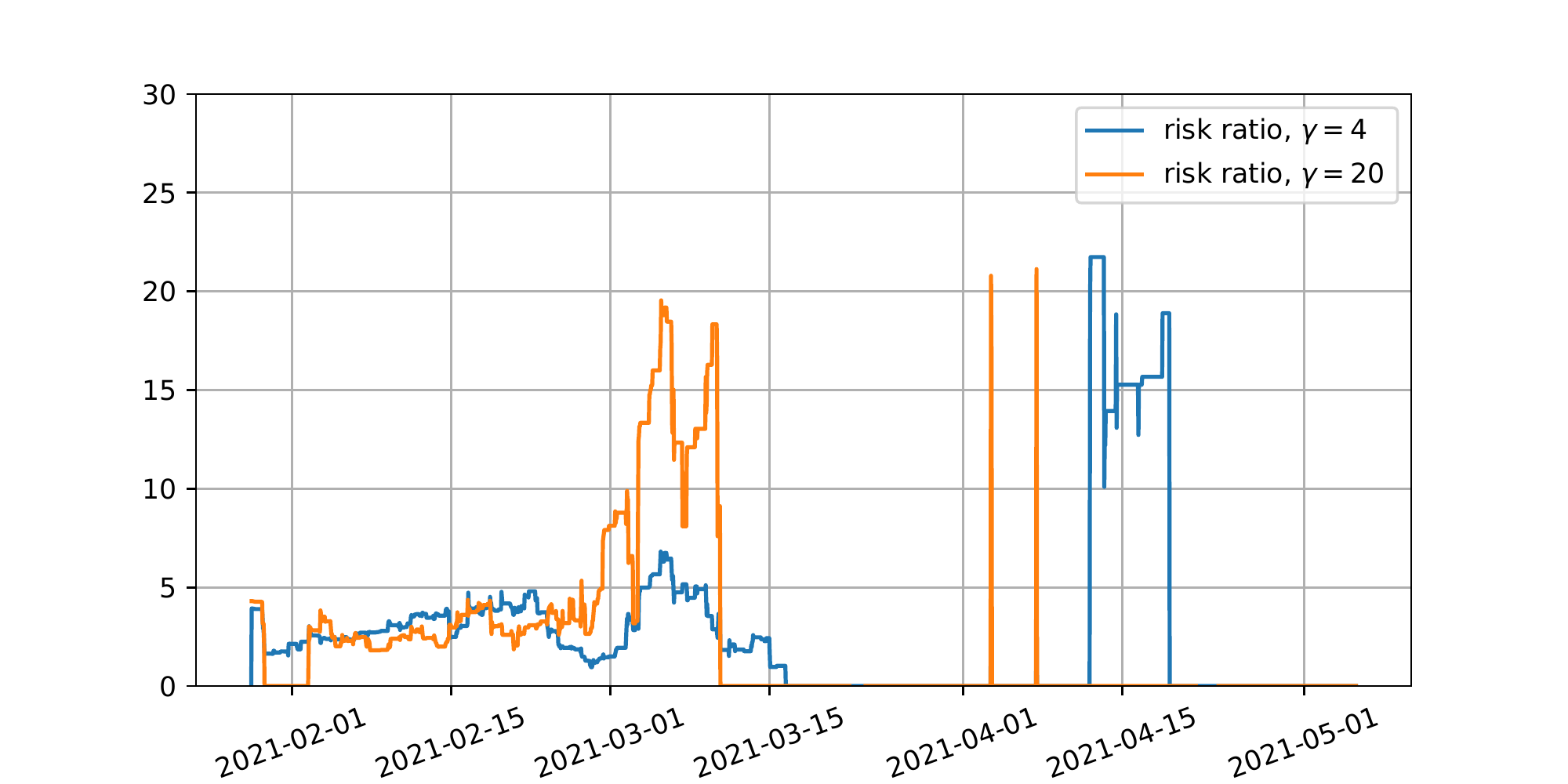}
\end{tabular}
\caption{\label{fig:rr} Risk ratio.  The risk ratio is defined as the odds of testing positive (in the following $\tau_p$ days) given a cumulative exposure score above $\gamma \in \{4, 20\}$, divided by the odds of testing positive (in the following $\tau_p$ days) given a cumulative exposure score below $\gamma \in \{4, 20\}$. $\tau_g = 7$, $\tau_s = 7$. Figure a, $\tau_p = 7$. Figure b $\tau_p = 28$.}
\end{figure*}

Fig. \ref{fig:TP_rate} and Fig. \ref{fig:rr} show the true positive rate, missed detection rate, and risk ratio when positive cases are predicted by an exposure score that exceeds a threshold $\gamma$ for a variety of algorithm parameters, \emph{as a function of time} during the study.   Note that Fig. \ref{fig:TP_rate} and Fig. \ref{fig:rr} show results on a \emph{per individual basis}.  In other words, the figures show the percent of the predicted \emph{individuals} that later test positive.  This is in contrast to Fig. \ref{fig:pp1} and Fig. \ref{fig:pp2}, which show the average number of contacts of individuals who test positive. 

\begin{figure}[htb]
\includegraphics[scale=0.41]{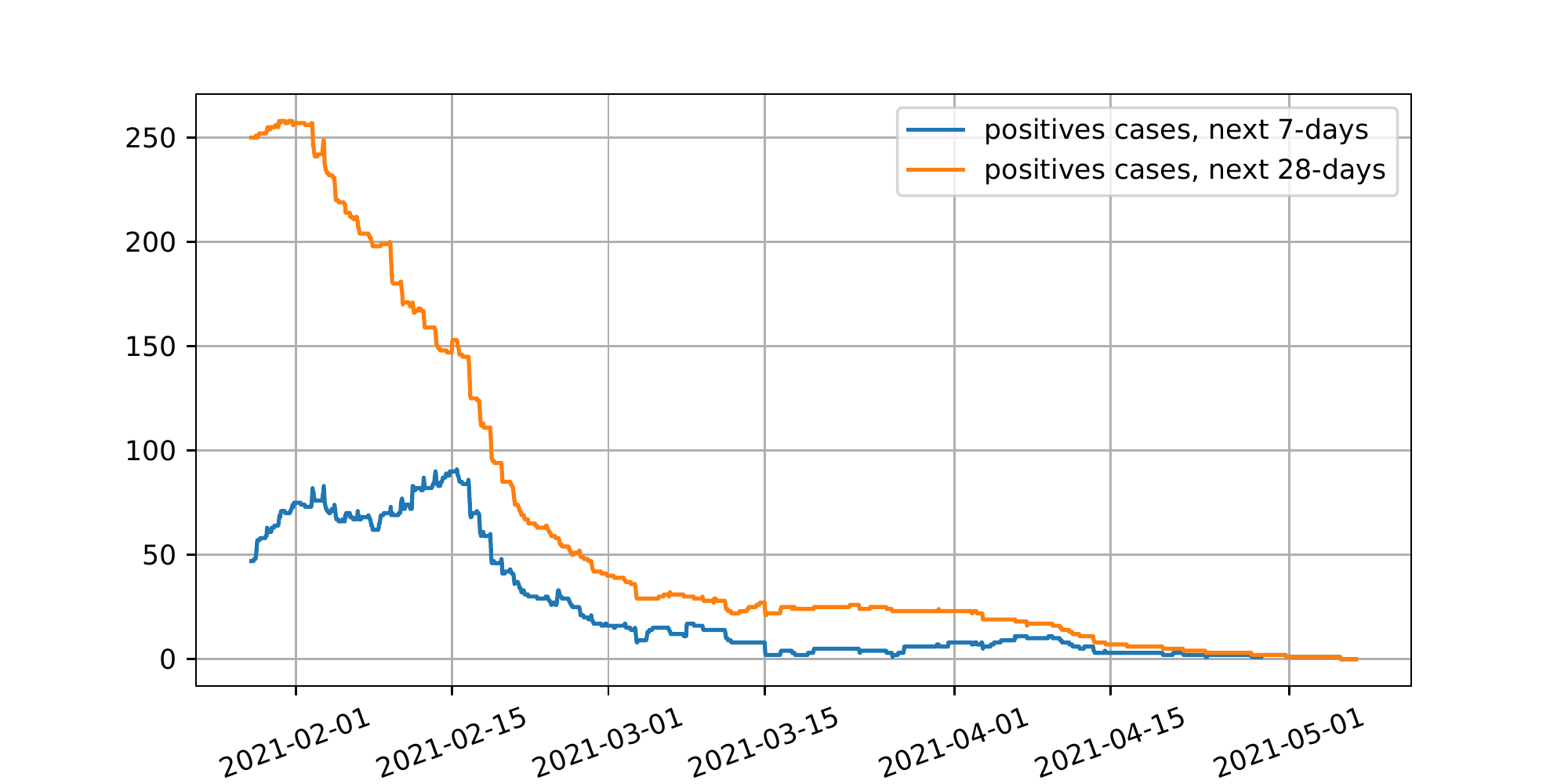}
\caption{ \label{fig:positive_case_count} Count of new positive cases in the $\tau_p = 7$ and $\tau_p = 28$ day period after date on the x-axis.}
\end{figure}

For reference, Fig. \ref{fig:positive_case_count} show the count of new positive cases in the $\tau_p \in \{7, 28\}$ days following the date on the x-axis. 

Lastly, we highlight the power of the dataset with a table that shows potential `high-spread' events. Table \ref{tab:high} shows the top (AP, hour) pairs when sorted by the highest percentage of positive users in the following 14 days.  We stress that the table does not imply that transmission occurred during these events.   

\begin{table*}[h!]
\begin{tabular}{|l|l|l|l|l|l|}
\hline
 AP & AP location & date and time & total users & positive users & positive users  \\ 
 & & & &  $\tau_p=7$ days & $\tau_p=14$ days \\ \hline
 A & residence hall & 2021-02-05, 20:00 - 21:00 & 14 & 8 & 10  \\ \hline
B & residence hall & 2021-02-10, 02:00 - 03:00 & 10 & 5 & 6  \\ \hline
C & residence hall & 2021-02-11, 12:00 - 13:00 & 13 & 7 & 7  \\ \hline
D & residence hall & 2021-02-11, 09:00 - 10:00 & 12 & 4 & 6  \\ \hline
E & residence hall & 2021-02-10, 13:00 - 14:00 & 11 & 3 & 5  \\ \hline
 F & dining hall & 2021-02-02, 19:00 - 20:00 & 499 & 17 & 25  \\ \hline
G & residence hall & 2021-02-04, 19:00 - 20:00 & 165 & 6 & 20  \\ \hline
H & dining hall & 2021-02-02, 19:00 - 20:00 & 361 & 12 & 20  \\ \hline
I & residence hall & 2021-02-07, 20:00 - 21:00 & 69 & 1 & 13  \\ \hline
J & residence hall & 2021-02-01, 19:00 - 20:00 & 53 & 8 & 12 \\ \hline
\end{tabular}
\caption{Potential high-spread events, February 2021.  Rows A-E, top (AP, hour) pairs with 10 or more users, sorted by highest percentage of users that test positive in the following $\tau_p = 14$ days.  Rows F-J, top (AP, hour) sorted by largest number of positive users in following $\tau_p =14$ days. Each AP is only listed once.  \label{tab:high}}
\end{table*}

\section{Discussion}
The results indicate that the use of network log data and the  AP-colocation algorithm has a positive predictive value of greater than 10\% (and  above 16\% under some parameter choices resulting in limited scale).  More specifically, at a scale of 1.7 contacts per positive case (on average), those returned contacts have greater than a 10\% chance of having a clinical positive COVID-19 result in the following 7 days.  At a higher scale, of 5.2 predicted contacts per positive case, the predicted contacts have a 5.0\% chance of a clinical positive in the following 7 days (see Table \ref{tab:res} and Fig. \ref{fig:pp2}). This is contrasted with the 0.79\% chance of a positive result in the next 7-days when contacts are selected at random from the ground truth dataset.  Comparing the 10\% predictive value of Alg. 1 to contacts chosen at random, the WiFi colocation contacts have a more than 12-fold increase in chances of a clinical positive test results in the following 7 days.

To account for exposure to multiple individuals that test positive, the \emph{exposure score} can be employed, providing a stronger predictive power.  Note that the false positive rate and false negative rates associated with a fixed threshold vary as the prevalence of the disease in the population varies.  Depending on the overall prevalence of the disease, users with an exposure score above a threshold exhibit a more than five-fold increase in odds of testing positive for COVID-19 over those with a score below the threshold (see Fig. \ref{fig:TP_rate}).

\section{Ethical Considerations}
With the collection of any network data there are natural privacy concerns, and there must be effort to balance the tradeoffs between these concerns and any potential benefit of such a system to the community. During a global pandemic, the potential benefits of better contact tracing are extremely high. Positive cases of \emph{reportable medical diseases} (such as COVID-19) must be reported to public health officials, who have broad authority to collect data related to contact tracing, whether those records are provided through oral history or electronically. Outside a pandemic, it is unlikely the risk/benefit to such a system is viable. During a pandemic, the specific risk/benefit depends on the prevalence of the disease, the severity of the disease, and the efficacy of other contact tracing approaches. All factors must be considered to establish if such a system is within an acceptable risk tolerance as it pertains to privacy, and we reference work that focuses on the privacy aspects of digital contact tracing \cite{hekmati2021contain, tang2020privacy, baumgartner2020mind, tahiliani2021privacy, tang2021another, ahmed2021dimy, hatamian2021privacy, holzapfel2020digital, ocheja2020quantifying, legendre2020contact, sanderson2021balancing}.

Graph datasets are attractive from a privacy perspective as their edges capture pair-wise relationships between individuals, and do not disclose the location (i.e, the AP or physical location).  Even if a contact graph is constructed from information that may be considered sensitive  (i.e, physical location), it can be stored and analyzed without any of the sensitive information.  Specific to our study, additional measures were taken to mitigate any disclosure of sensitive information.  All identifiers -- MAC addresses, authentication information, AP identifiers -- were anonymized via a one way hash before analysis was completed, and the authors of the study never had access to plain text user identifiers.

Lastly, this work has been determined by the Minimal Risk IRB at the University of Wisconsin as not research involving human subjects as defined by the United States Department of Health and Human Services (DHHS) and Unites States Food and Drug Administration (FDA).  We reference the Menlo Report~\cite{Dittrich12} as it establishes ethical principles and provides context for these principles in computing and communication research.  The Menlo Report provides guidance with respect to Institutional Review Board (IRB) and self evaluation.

\section{Related Work}
Digital contact tracing has received a surge of attention since the start of the COVID-19 pandemic.  Many of the initial approaches required installation of a mobile application for data collection and rely on location services (i.e, GPS) or proximity sensing using technologies such as Bluetooth low energy (BLE).  Similar to app-based approaches, operating system level approaches use the operating system (OS) to collect data after a user `opts-in', but in both cases, data is collected on the client device.   Notably, in a joint venture, Google and Apple developed and released contact tracing technology known as GAEN (Google/Apple Exposure Notification) \cite{apple_google}, which relies on a BLE protocol to sense surrounding devices.  Both app-based and OS based technologies have been adopted by governments with varying degrees of success.  In some instances, where state level governments \emph{highly recommended} contact tracing applications, installation was noted to be under 3\% of adults~\cite{time_nevada}. Furthermore, while the relatively short range of Bluetooth would initially seem an advantage in identifying close contacts, growing evidence of aerosol transmission of SARS-CoV-2 at distances of up to 60 feet~\cite{Bazante2018995118}, which more closely matches indoor WiFi transmission range, suggests that tools based on WiFi would be a useful addition to the contact tracing tool set.

In contrast to approaches based on client-side data collection, network-side contact tracing, such as the approach proposed in this paper, can be implemented without the installation of an app.  Closely related to the work in this paper is that of \cite{trivedi2021wifitrace, zakaria2020analyzing}. In \cite{trivedi2021wifitrace}, the authors propose use of WiFi network association logs gathered by enterprise networks to create a graph data structure which can be used for contact tracing.  The approach (including Alg. 1 of \cite{trivedi2021wifitrace}) is similar to proposed co-location graph algorithms for other applications, including \cite{malloy2017internet, funkhouser2018device}.  The authors of \cite{trivedi2021wifitrace} implement their approach and demonstrate its efficacy with WiFi datasets but simulated disease outbreak data, as they do not have access to ground truth COVID test results. Other campus-wide studies related to the COVID-19 pandemic using network log data include 
\cite{zhang2021wlan} which studies super-spreader events using data collected on a university campus. Many additional papers focused on cellular or WiFi networks have surfaced during the COVID-19 pandemic \cite{dmitrienko2020proximity, tu2021epidemic, mcheick2021d2d, zagatti2021large, yoo2020bim,  manavi2020review, liu2021wibeacon, yi2021cellular, monroe2021location, giustiniano20215g, oikonomidis2021role, zhao2020accuracy, abowd2020using, thakare2021p, bressandata, nguyen2020epidemic, zang2021building, li2021vcontact} which do not include ground truth datasets. Overview articles that explore WiFi technologies include \cite{petrovic2021iot, sahraouitraceme, sun2021mitigating, basheeruddinasdaq2021wireless, braithwaiteautomated, braithwaite2020automated, roy2020efficient}.

Another closely related work is the approach proposed in \cite{malloy2020digital}, which relies on data collected by entities in the digital advertising ecosystem, and \emph{IP-colocation} (as opposed to \emph{AP-colocation}). The techniques and construction of the co-location graph are similar.  IP-colocation and AP-colocation are likely to exhibit significant overlap. In many homes and small businesses, WiFi users spend significant time in close proximity and share a single public IP address through NAT. Enterprise networks, such as the WiFi network in this study, have more flexibility in managing their IP address space with policy options ranging from assigning each device a unique public address to arbitrarily distributing devices across a pool of public addresses through NAT. 

Finally, network-side contact tracing has some overlap with the problem of WiFi localization, particularly from the vantage point of locating transmitters within a network. Approaches based on received signal strength~\cite{bahl2000radar}, angle of arrival~\cite{joshi2013pinpoint}, and time of flight~\cite{giustiniano2011caesar} all have the potential to improve contact tracing accuracy by providing an improved measure of physical proximity. We speculate that narrowing the spatial resolution to the room level is likely to be of epidemiological value. 
Additionally, if the wireless network facilitates tracking of user devices with finer granularity than association and disassociation events, it would enable a more precise calculation of AP colocation duration despite devices roaming without explicit disassociation messages.

\section{Summary}
Our results suggest that network-side contact tracing can help identify individuals at increased risk of disease plausibly through exposure in a shared environment. In contexts where available, e.g. academic and corporate campuses, Wi-Fi networks can be effectively leveraged to help control disease spread.  The approaches discussed here can be extrapolated to other settings in which network log data is collected. 

While the confidence scores defined by algorithm~\ref{alg:graph} are predictive of positive cases, determining actionable and effective criteria to implement interventions is a sizable outstanding effort. In the future, more work must be done before such a system is viable, in particular working with health and administrative officials to determine if notification of possible exposure to SARS-CoV-19 (or other communicable diseases) is warranted.


\section*{Appendix} \label{sec:extended_results} 

\begin{figure*}[h]\centering
  \begin{tabular}{cc}
   \includegraphics[scale=0.41]{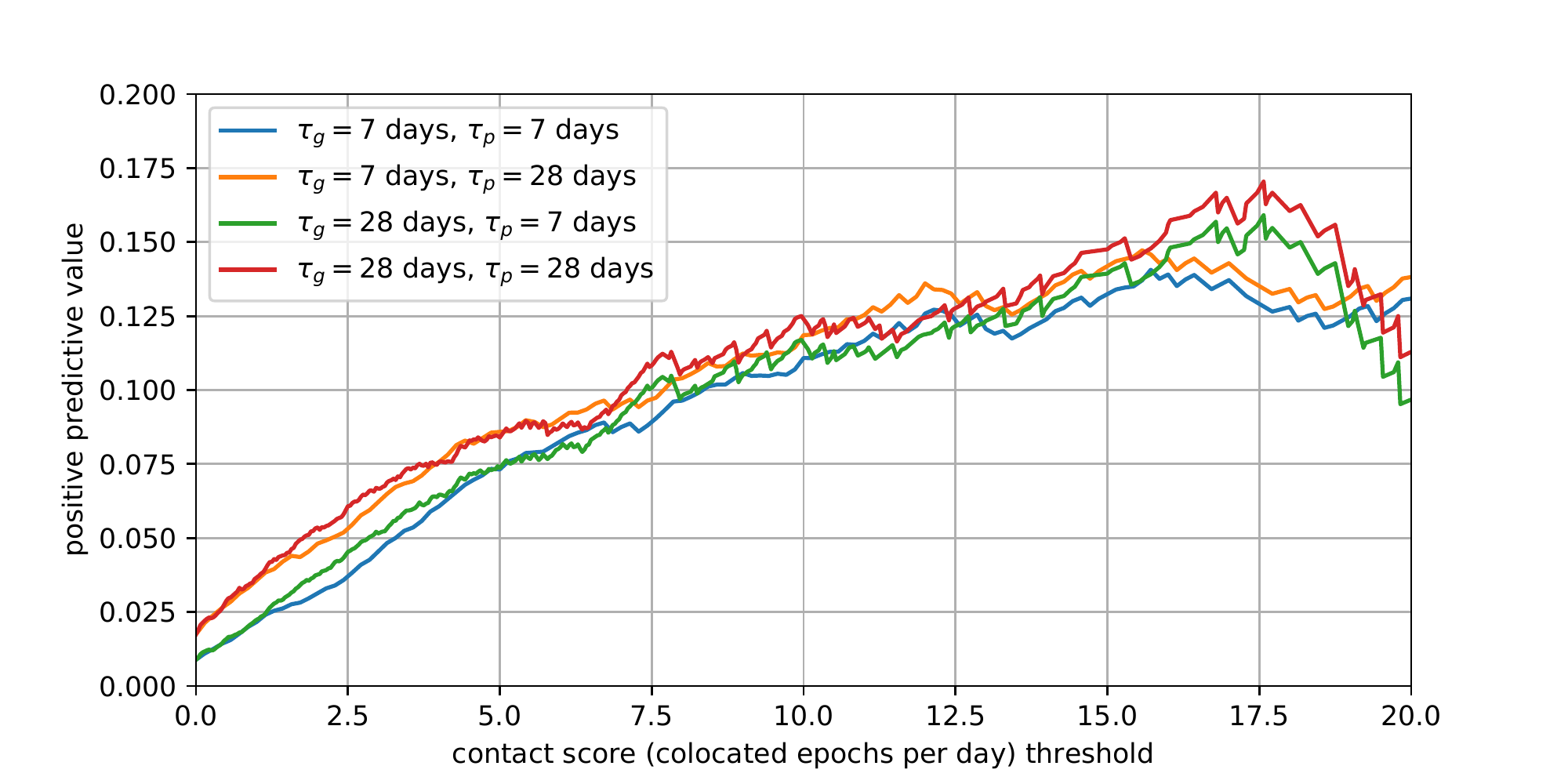} & \includegraphics[scale=0.41]{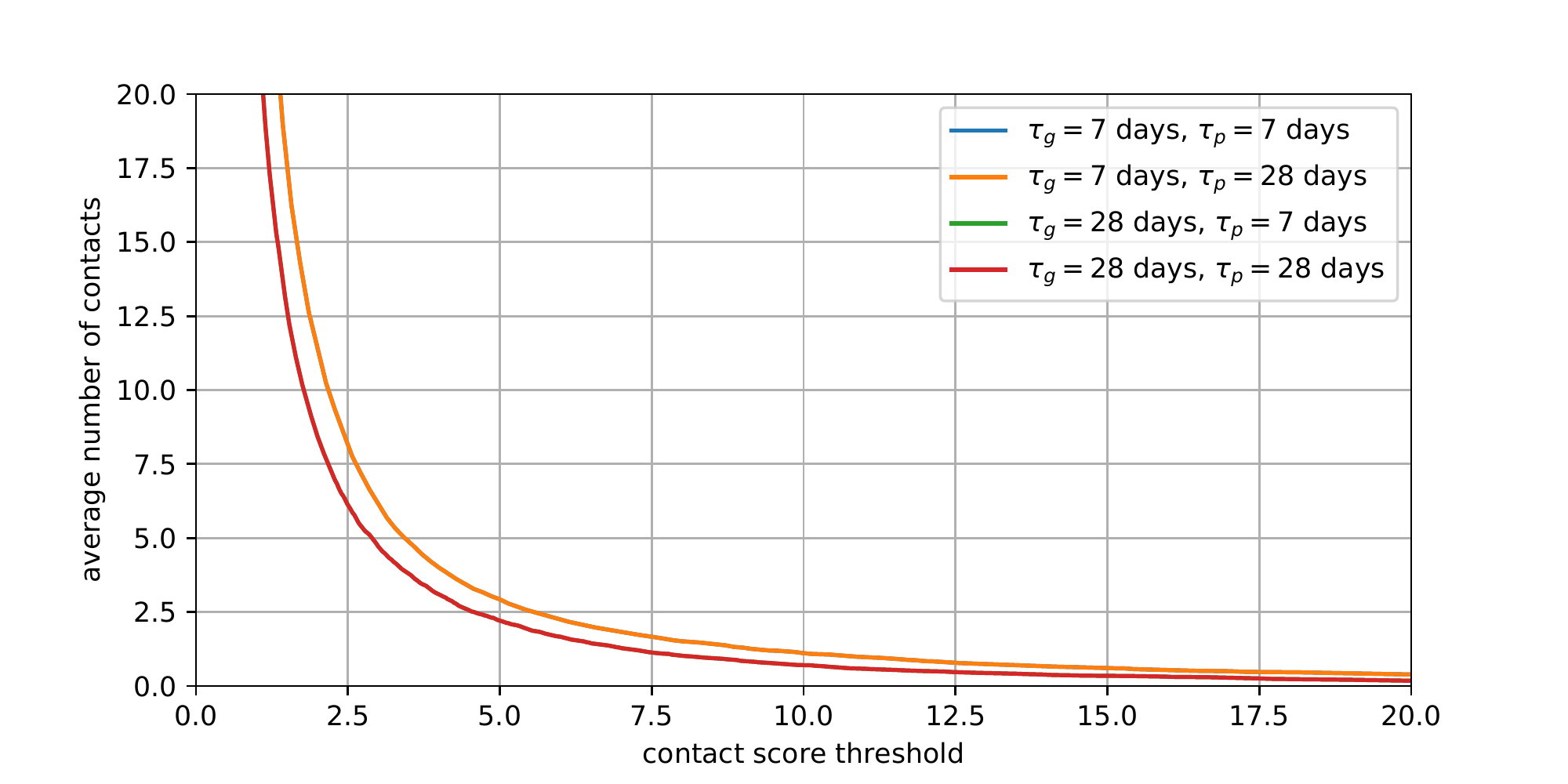} \\
   (a) & (b) \\
   \end{tabular}
  \caption{Figure a.  $\alpha = 0$. Percent of predicted contacts that test positive in the $\tau_p$-days ($\tau_p \in \{7,28\}$) after a positive case as a function of the threshold $\gamma$, denoted PPV. Contact graph with $\tau_g \in \{7,28\}$.  Figure b.  
  Average number of predicted contacts as a function of contact score $\gamma$ for $\tau_g \in \{7,28\}$ and $\tau_p \in \{7,28\}$. Note that the plots do not depend on $\tau_p$; two traces are obscured.}
\end{figure*}

\subsection*{Ground Truth} \label{sec:ground_truth} 
At the start of the Spring 2021 semester, students returned to campus housing facilities but required frequent testing with rapid saliva-based COVID-19 tests. Students who were residents of university housing and tested positive were required to move into one of five designated isolation dormitories or self-isolate in off-campus housing. Access to the isolation facilities was controlled by the university administration, which closely monitored and recorded the daily occupancy of each dormitory. Since all of the campus dormitories including the isolation dorms are also covered by campus WiFi infrastructure, we hypothesized that the WiFi logs would offer insight into devices that moved from regular dorms to isolation dorms without revealing personally identifiable information about the device owner.  To this end, we assessed the validity of our hypothesis by comparing the number of devices detected in isolation dorms with the true count of admitted residents supplied by the university.

While visualizing device activity over time, we quickly realized that time of day has a large impact on observed WiFi activity. During busy hours of the day, we would overestimate the number of residents by a factor of two or more because of the presence of staff. A prior study of device type and user behavior~\cite{trivedi2020empirical} suggests differences in the types of devices most likely to be found at varying hours of the day. In the early hours of the morning when dormitory residents are likely to be present and sleeping and staff or visitors are likely absent, we expect any devices still connected to the WiFi network are most likely to be the personal phones belonging to residents. Whereas phones may maintain a network connection in order to receive updates, most other devices such as laptops and entertainment devices are likely to be in a low power, non-transmitting state. If true, this tendency could not only assist us inferring residence but could also help identify the devices most likely to be carried with the person for contact tracing.

\begin{figure}
\includegraphics[width=0.39\textwidth]{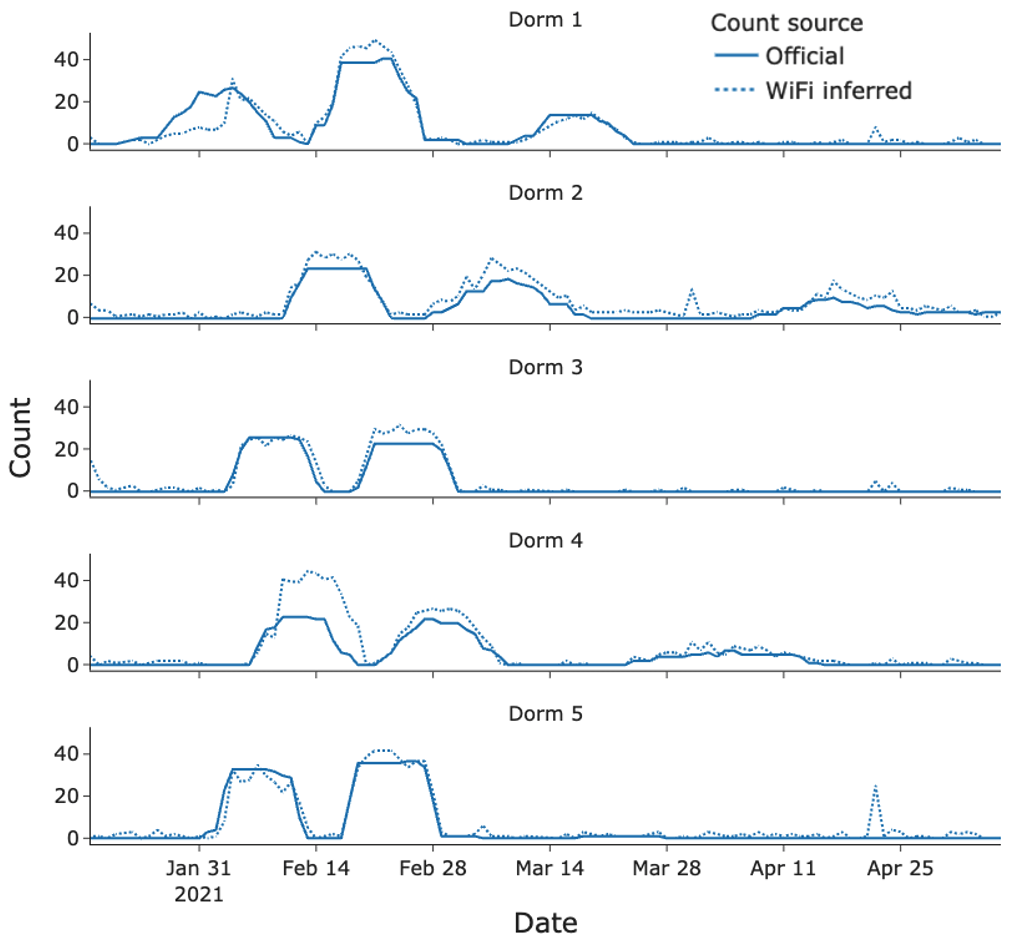}
\caption{Device counts detected in the early morning hour of 4:00-5:00 AM compared with true number of residents in five isolation dorms.}
\label{fig:morning-counts}
\end{figure}

To test the feasibility of estimating building occupancy, we compare the number of unique devices (anonymized MAC addresses) detected during different times of the day with the daily resident counts from the campus health authorities. We frame this as an optimization problem of choosing the best hour of the day to use WiFi device counts to predict the number of residents in a dormitory. More specifically, we count the number of devices detected in the five isolation dorms during a given one hour window for each day over the course of the study and compute the mean squared error (MSE) against the true counts in the respective dorms. We repeat this for each of the 24 hours of the day and find that the one hour window of 4:00-5:00 AM minimizes the prediction error, as shown in figure~\ref{fig:mse-residents}. We find that the resulting device counts are a good approximation of the true number of residents. Figure~\ref{fig:morning-counts} shows the five isolation dorms with inferred and true resident counts. Compared to the general problem of predicting a building's occupancy from WiFi activity, highly predictable device and user behavior make dorm residency a relatively easier problem.

\subsection*{Inferring Dorm Residents}

After finding a correlation between the number of devices detected in the early morning and the number of isolation dorm residents, we turn our attention to identifying devices and their corresponding owners who reside in any of the campus dormitories. Recall that for the purpose of evaluating the effectiveness of our contact tracing approach, we must limit our population to the users who live in campus housing because this is the subset of users for whom we can infer COVID-19 cases based on presence in the isolation dorms. Although we do not have ground truth numbers for Spring semester residents, we take the reported capacity from university housing web pages as an estimate of the true numbers. We find that simply considering any device detected during the hour of 4:00-5:00 AM as belonging to a resident leads to a large overcount, perhaps explained in part by visitors, so it is necessary to filter the more probable residents.

\begin{figure}
\centering
\includegraphics[width=0.41\textwidth]{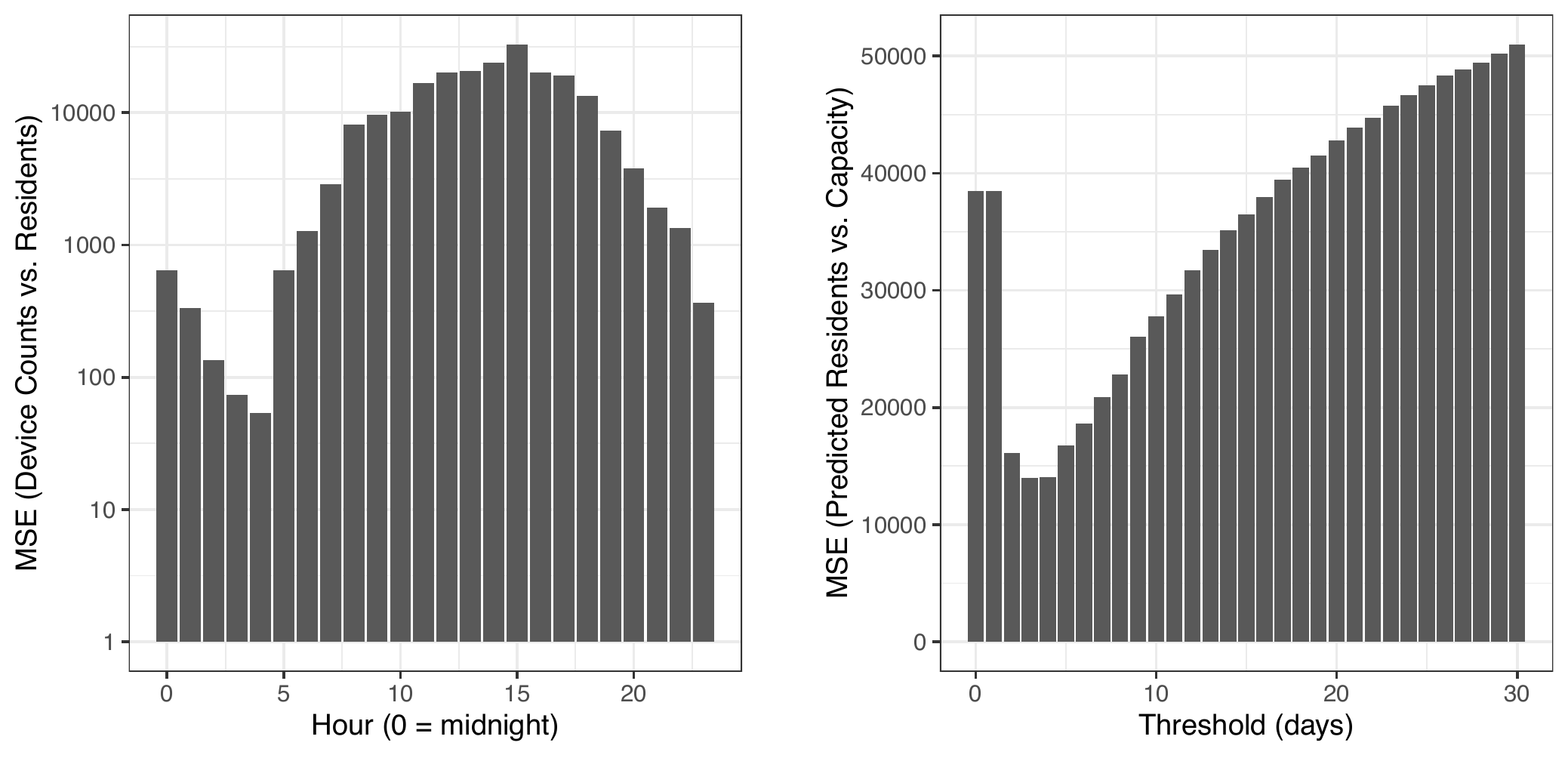}
\caption{
\label{fig:mse-residents}MSE of device count at different times of the day as a predictor of isolation dorm residents. (right).  
MSE of inferred resident counts compared to reported building capacities for 19 dormitories.
} 
\end{figure}

This time operating over the range of user IDs, we count the number of days in which any device belonging to a user is detected in a building during the hour of 4:00-5:00 AM. If multiple devices belonging to the same user are detected on the same morning, we only attribute one detection to that building. If the number of times the user is detected in a given building meets or exceeds a threshold, $\tau_r$, then we consider the user a probable resident of that building. For a given value of $\tau_r$ we count the number of inferred residents in each building, which strictly decreases as $\tau_r$ increases. Iterating over $\tau_r = 0, 1, 2, ..., 30$, we find that $\tau_r = 3$ minimizes the MSE against reported dorm capacity numbers, as seen in figure~\ref{fig:mse-residents}. We take the users meeting this threshold as the population of dorm residents for the remainder of our study and disregard non-residents. Although we are able to compute AP colocation metrics for non-resident devices and perform digital contact tracing, we are only able to evaluate efficacy for contact tracing in the COVID-19 pandemic on the subset of devices that belong to residents.

We briefly note two potential issues with this approach. First, this choice of threshold is still expected to result in overestimation because distancing measures and more students choosing to live off campus likely meant that dorms were under capacity during the semester. However, we felt it important not to falsely exclude too many users from our evaluation, and furthermore, overestimating the population of residents is unlikely to bias the results in our favor. Second, under this approach a small number of users may be considered as probable residents of more than one building. Since our goal is to identify the set of dorm residents, we do not consider that a problem, but if we needed to infer a user's building of residence, some refinements could be made such as selecting the building with the highest detection count.

\subsection*{Inferring Isolation Residents}

\begin{figure}
\centering
\includegraphics[width=0.31\textwidth]{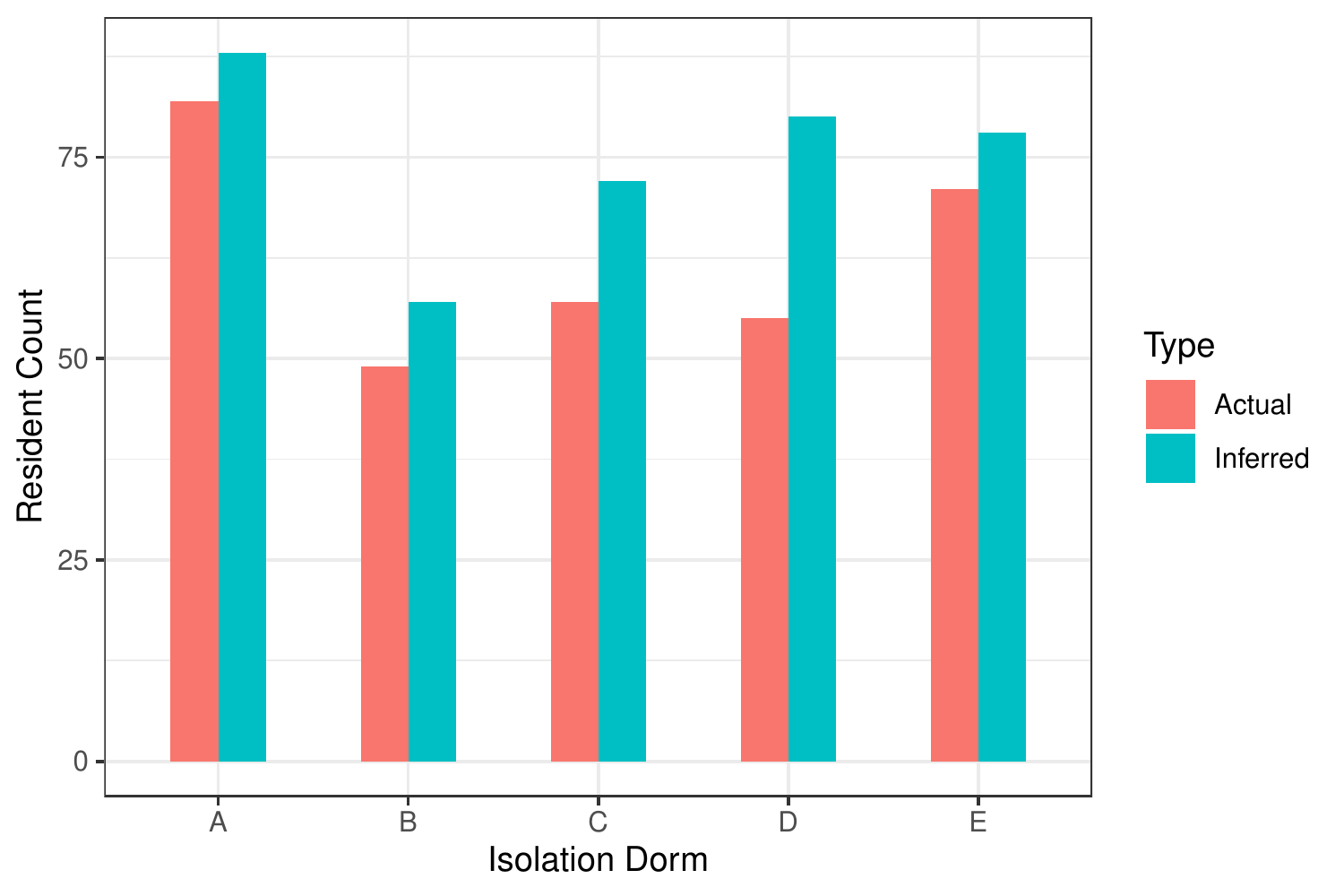}
\caption{\label{fig:isolation-counts} Number of inferred users in each isolation dorm compared with true counts.}
\end{figure}

Finally, we apply a similar approach from the previous section to identify dorm residents who temporarily move into one of the five isolation dorms. Using the same threshold, $\tau_r = 3$, we scan the association logs for users whose device(s) appears in an isolation dorm for at least $\tau_r$ mornings and label these as positive cases.  Starting from the first time one of the user's devices is detected in isolation, we find the longest uninterrupted length of time until any of the user's devices is detected in a different building, and consider that the user's length of stay in isolation. Figure~\ref{fig:isolation-counts} shows the number of users inferred as residents in each isolation dorm compared to the ground truth data for the semester, which we received from campus health authorities.

\section*{Acknowledgements}
Suman Banerjee and Lance Hartung are supported in part through the following US National Science Foundation grants:
CNS-1719336, CNS-1647152, CNS-1629833, and CNS-2003129 and an award from the US Department of Commerce with award number 70NANB21H043.
\bibliography{wifi_contact}


\begin{thebibliography}{52}


\ifx \showCODEN    \undefined \def \showCODEN     #1{\unskip}     \fi
\ifx \showDOI      \undefined \def \showDOI       #1{#1}\fi
\ifx \showISBNx    \undefined \def \showISBNx     #1{\unskip}     \fi
\ifx \showISBNxiii \undefined \def \showISBNxiii  #1{\unskip}     \fi
\ifx \showISSN     \undefined \def \showISSN      #1{\unskip}     \fi
\ifx \showLCCN     \undefined \def \showLCCN      #1{\unskip}     \fi
\ifx \shownote     \undefined \def \shownote      #1{#1}          \fi
\ifx \showarticletitle \undefined \def \showarticletitle #1{#1}   \fi
\ifx \showURL      \undefined \def \showURL       {\relax}        \fi
\providecommand\bibfield[2]{#2}
\providecommand\bibinfo[2]{#2}
\providecommand\natexlab[1]{#1}
\providecommand\showeprint[2][]{arXiv:#2}

\bibitem[\protect\citeauthoryear{Abowd}{Abowd}{2020}]%
        {abowd2020using}
\bibfield{author}{\bibinfo{person}{Gregory~D Abowd}.}
  \bibinfo{year}{2020}\natexlab{}.
\newblock \showarticletitle{Using digital technologies to support pandemic
  response on campus: A case study in the opportunities and challenges of
  WiFi}.
\newblock  (\bibinfo{year}{2020}).
\newblock


\bibitem[\protect\citeauthoryear{Ahmed, Michelin, Xue, Putra, Ruj, Kanhere, and
  Jha}{Ahmed et~al\mbox{.}}{2021}]%
        {ahmed2021dimy}
\bibfield{author}{\bibinfo{person}{Nadeem Ahmed}, \bibinfo{person}{Regio~A
  Michelin}, \bibinfo{person}{Wanli Xue}, \bibinfo{person}{Guntur~Dharma
  Putra}, \bibinfo{person}{Sushmita Ruj}, \bibinfo{person}{Salil~S Kanhere},
  {and} \bibinfo{person}{Sanjay Jha}.} \bibinfo{year}{2021}\natexlab{}.
\newblock \showarticletitle{DIMY: Enabling privacy-preserving contact tracing}.
\newblock \bibinfo{journal}{\emph{arXiv preprint arXiv:2103.05873}}
  (\bibinfo{year}{2021}).
\newblock


\bibitem[\protect\citeauthoryear{Bahl and Padmanabhan}{Bahl and
  Padmanabhan}{2000}]%
        {bahl2000radar}
\bibfield{author}{\bibinfo{person}{Victor Bahl} {and} \bibinfo{person}{Venkat
  Padmanabhan}.} \bibinfo{year}{2000}\natexlab{}.
\newblock \showarticletitle{RADAR: An In-Building RF-based User Location and
  Tracking System}. In \bibinfo{booktitle}{\emph{Proceedings of IEEE INFOCOM
  2000}}. \bibinfo{publisher}{Institute of Electrical and Electronics
  Engineers, Inc.}
\newblock
\urldef\tempurl%
\url{https://www.microsoft.com/en-us/research/publication/radar-an-in-building-rf-based-user-location-and-tracking-system/}
\showURL{%
\tempurl}
\newblock
\shownote{ACM SIGMOBILE Test-of-Time Paper Award, 2016.}


\bibitem[\protect\citeauthoryear{Basheeruddin~Asdaq, Naveen, Gunturu,
  Pamayyagari, Abdullah, Sreeharsha, Imran, Alsalman, Al~Hawaj, and
  Alsubaie}{Basheeruddin~Asdaq et~al\mbox{.}}{2021}]%
        {basheeruddinasdaq2021wireless}
\bibfield{author}{\bibinfo{person}{Syed~Mohammed Basheeruddin~Asdaq},
  \bibinfo{person}{N~Raghavendra Naveen}, \bibinfo{person}{Lakshmi~Narasimha
  Gunturu}, \bibinfo{person}{Kalpana Pamayyagari}, \bibinfo{person}{Ibrahim
  Abdullah}, \bibinfo{person}{Nagaraja Sreeharsha}, \bibinfo{person}{Mohd
  Imran}, \bibinfo{person}{Abdulkhaliq~J Alsalman}, \bibinfo{person}{Maitham~A
  Al~Hawaj}, {and} \bibinfo{person}{Abdullah~A Alsubaie}.}
  \bibinfo{year}{2021}\natexlab{}.
\newblock \showarticletitle{Wireless Networking-Driven Healthcare Approaches in
  Combating COVID-19}.
\newblock \bibinfo{journal}{\emph{BioMed Research International}}
  \bibinfo{volume}{2021} (\bibinfo{year}{2021}).
\newblock


\bibitem[\protect\citeauthoryear{Baumg{\"a}rtner, Dmitrienko, Freisleben,
  Gruler, H{\"o}chst, K{\"u}hlberg, Mezini, Mitev, Miettinen, Muhamedagic,
  et~al\mbox{.}}{Baumg{\"a}rtner et~al\mbox{.}}{2020}]%
        {baumgartner2020mind}
\bibfield{author}{\bibinfo{person}{Lars Baumg{\"a}rtner},
  \bibinfo{person}{Alexandra Dmitrienko}, \bibinfo{person}{Bernd Freisleben},
  \bibinfo{person}{Alexander Gruler}, \bibinfo{person}{Jonas H{\"o}chst},
  \bibinfo{person}{Joshua K{\"u}hlberg}, \bibinfo{person}{Mira Mezini},
  \bibinfo{person}{Richard Mitev}, \bibinfo{person}{Markus Miettinen},
  \bibinfo{person}{Anel Muhamedagic}, {et~al\mbox{.}}}
  \bibinfo{year}{2020}\natexlab{}.
\newblock \showarticletitle{Mind the GAP: Security \& privacy risks of contact
  tracing apps}. In \bibinfo{booktitle}{\emph{2020 IEEE 19th International
  Conference on Trust, Security and Privacy in Computing and Communications
  (TrustCom)}}. IEEE, \bibinfo{pages}{458--467}.
\newblock


\bibitem[\protect\citeauthoryear{Bazant and Bush}{Bazant and Bush}{2021}]%
        {Bazante2018995118}
\bibfield{author}{\bibinfo{person}{Martin~Z. Bazant} {and}
  \bibinfo{person}{John W.~M. Bush}.} \bibinfo{year}{2021}\natexlab{}.
\newblock \showarticletitle{A guideline to limit indoor airborne transmission
  of COVID-19}.
\newblock \bibinfo{journal}{\emph{Proceedings of the National Academy of
  Sciences}} \bibinfo{volume}{118}, \bibinfo{number}{17}
  (\bibinfo{year}{2021}).
\newblock
\showISSN{0027-8424}
\urldef\tempurl%
\url{https://doi.org/10.1073/pnas.2018995118}
\showDOI{\tempurl}
\showeprint{https://www.pnas.org/content/118/17/e2018995118.full.pdf}


\bibitem[\protect\citeauthoryear{Braithwaite, Callender, Bullock, and
  Aldridge}{Braithwaite et~al\mbox{.}}{2020b}]%
        {braithwaite2020automated}
\bibfield{author}{\bibinfo{person}{Isobel Braithwaite}, \bibinfo{person}{Tom
  Callender}, \bibinfo{person}{Miriam Bullock}, {and} \bibinfo{person}{Robert~W
  Aldridge}.} \bibinfo{year}{2020}\natexlab{b}.
\newblock \showarticletitle{Automated and partially-automated contact tracing:
  a rapid systematic review to inform the control of COVID-19}.
\newblock \bibinfo{journal}{\emph{medRxiv}} (\bibinfo{year}{2020}).
\newblock


\bibitem[\protect\citeauthoryear{Braithwaite, Callender, Bullock, Aldridge, and
  Braithwaite}{Braithwaite et~al\mbox{.}}{2020a}]%
        {braithwaiteautomated}
\bibfield{author}{\bibinfo{person}{Isobel Braithwaite}, \bibinfo{person}{Tom
  Callender}, \bibinfo{person}{Miriam Bullock}, \bibinfo{person}{Robert~W
  Aldridge}, {and} \bibinfo{person}{Isobel Braithwaite}.}
  \bibinfo{year}{2020}\natexlab{a}.
\newblock \showarticletitle{Automated and partially-automated contact tracing:
  a rapid systematic review to inform the control of COVID-19}.
\newblock \bibinfo{journal}{\emph{medRxiv}} (\bibinfo{year}{2020}).
\newblock


\bibitem[\protect\citeauthoryear{Bressan}{Bressan}{2021}]%
        {bressandata}
\bibfield{author}{\bibinfo{person}{St{\'e}phane Bressan}.}
  \bibinfo{year}{2021}\natexlab{}.
\newblock \showarticletitle{A Data Warehouse of Wi-Fi Sessions for Contact
  Tracing and Outbreak Investigation}.
\newblock \bibinfo{journal}{\emph{Transactions on Large-Scale Data-and
  Knowledge-Centered Systems XLVIII: Special Issue In Memory of Univ. Prof. Dr.
  Roland Wagner}} (\bibinfo{year}{2021}), \bibinfo{pages}{85}.
\newblock


\bibitem[\protect\citeauthoryear{Dittrich and Kenneally}{Dittrich and
  Kenneally}{2012}]%
        {Dittrich12}
\bibfield{author}{\bibinfo{person}{D. Dittrich} {and} \bibinfo{person}{E.
  Kenneally}.} \bibinfo{year}{2012}\natexlab{}.
\newblock \showarticletitle{{The Menlo Report: Ethical Principles Guiding
  Information and Communication Technology Research}}. In
  \bibinfo{booktitle}{\emph{US Department of Homeland Security Tech Report}}.
\newblock


\bibitem[\protect\citeauthoryear{Dmitrienko, Singh, Erichsen, and
  Raskar}{Dmitrienko et~al\mbox{.}}{2020}]%
        {dmitrienko2020proximity}
\bibfield{author}{\bibinfo{person}{Mikhail Dmitrienko},
  \bibinfo{person}{Abhishek Singh}, \bibinfo{person}{Patrick Erichsen}, {and}
  \bibinfo{person}{Ramesh Raskar}.} \bibinfo{year}{2020}\natexlab{}.
\newblock \showarticletitle{Proximity Inference with Wifi-Colocation during the
  COVID-19 Pandemic}.
\newblock \bibinfo{journal}{\emph{arXiv preprint arXiv:2009.12699}}
  (\bibinfo{year}{2020}).
\newblock


\bibitem[\protect\citeauthoryear{Funkhouser, Malloy, Alp, Poon, and
  Barford}{Funkhouser et~al\mbox{.}}{2018}]%
        {funkhouser2018device}
\bibfield{author}{\bibinfo{person}{Keith Funkhouser}, \bibinfo{person}{Matthew
  Malloy}, \bibinfo{person}{Enis~Ceyhun Alp}, \bibinfo{person}{Phillip Poon},
  {and} \bibinfo{person}{Paul Barford}.} \bibinfo{year}{2018}\natexlab{}.
\newblock \showarticletitle{Device Graphing by Example}. In
  \bibinfo{booktitle}{\emph{Proceedings of the 24th ACM SIGKDD International
  Conference on Knowledge Discovery \& Data Mining}}.
  \bibinfo{pages}{273--282}.
\newblock


\bibitem[\protect\citeauthoryear{Garza}{Garza}{2020a}]%
        {apple_google}
\bibfield{author}{\bibinfo{person}{Alejandro De~La Garza}.}
  \bibinfo{year}{2020}\natexlab{a}.
\newblock \showarticletitle{Apple and Google Partner on {COVID-19} Contact
  Tracing Technology}.
\newblock
  \bibinfo{journal}{\emph{https://www.apple.com/newsroom/2020/04/apple-and-google-partner-on-covid-19-contact-tracing-technology/}}
  (\bibinfo{year}{2020}).
\newblock


\bibitem[\protect\citeauthoryear{Garza}{Garza}{2020b}]%
        {time_nevada}
\bibfield{author}{\bibinfo{person}{Alejandro De~La Garza}.}
  \bibinfo{year}{2020}\natexlab{b}.
\newblock \showarticletitle{Contact Tracing Apps Were Big Tech's Best Idea for
  Fighting COVID-19. Why Haven't They Helped?}
\newblock \bibinfo{journal}{\emph{Time Magazine}} (\bibinfo{date}{November}
  \bibinfo{year}{2020}).
\newblock


\bibitem[\protect\citeauthoryear{Giustiniano, Bianchi, Conti, Bartoletti, and
  Melazzi}{Giustiniano et~al\mbox{.}}{2021}]%
        {giustiniano20215g}
\bibfield{author}{\bibinfo{person}{Domenico Giustiniano},
  \bibinfo{person}{Giuseppe Bianchi}, \bibinfo{person}{Andrea Conti},
  \bibinfo{person}{Stefania Bartoletti}, {and} \bibinfo{person}{Nicola~Blefari
  Melazzi}.} \bibinfo{year}{2021}\natexlab{}.
\newblock \showarticletitle{5G and Beyond for Contact Tracing}.
\newblock \bibinfo{journal}{\emph{IEEE Communications Magazine}}
  \bibinfo{volume}{59}, \bibinfo{number}{9} (\bibinfo{year}{2021}),
  \bibinfo{pages}{36--41}.
\newblock


\bibitem[\protect\citeauthoryear{Giustiniano and Mangold}{Giustiniano and
  Mangold}{2011}]%
        {giustiniano2011caesar}
\bibfield{author}{\bibinfo{person}{Domenico Giustiniano} {and}
  \bibinfo{person}{Stefan Mangold}.} \bibinfo{year}{2011}\natexlab{}.
\newblock \showarticletitle{CAESAR: Carrier Sense-Based Ranging in
  off-the-Shelf 802.11 Wireless LAN}. In \bibinfo{booktitle}{\emph{Proceedings
  of the Seventh COnference on Emerging Networking EXperiments and
  Technologies}} (Tokyo, Japan) \emph{(\bibinfo{series}{CoNEXT '11})}.
  \bibinfo{publisher}{Association for Computing Machinery},
  \bibinfo{address}{New York, NY, USA}, Article \bibinfo{articleno}{10},
  \bibinfo{numpages}{12}~pages.
\newblock
\showISBNx{9781450310413}
\urldef\tempurl%
\url{https://doi.org/10.1145/2079296.2079306}
\showDOI{\tempurl}


\bibitem[\protect\citeauthoryear{Hatamian, Wairimu, Momen, and
  Fritsch}{Hatamian et~al\mbox{.}}{2021}]%
        {hatamian2021privacy}
\bibfield{author}{\bibinfo{person}{Majid Hatamian}, \bibinfo{person}{Samuel
  Wairimu}, \bibinfo{person}{Nurul Momen}, {and} \bibinfo{person}{Lothar
  Fritsch}.} \bibinfo{year}{2021}\natexlab{}.
\newblock \showarticletitle{A privacy and security analysis of early-deployed
  COVID-19 contact tracing Android apps}.
\newblock \bibinfo{journal}{\emph{Empirical Software Engineering}}
  \bibinfo{volume}{26}, \bibinfo{number}{3} (\bibinfo{year}{2021}),
  \bibinfo{pages}{1--51}.
\newblock


\bibitem[\protect\citeauthoryear{Hekmati, Ramachandran, and
  Krishnamachari}{Hekmati et~al\mbox{.}}{2021}]%
        {hekmati2021contain}
\bibfield{author}{\bibinfo{person}{Arvin Hekmati}, \bibinfo{person}{Gowri
  Ramachandran}, {and} \bibinfo{person}{Bhaskar Krishnamachari}.}
  \bibinfo{year}{2021}\natexlab{}.
\newblock \showarticletitle{CONTAIN: Privacy-oriented contact tracing protocols
  for epidemics}. In \bibinfo{booktitle}{\emph{2021 IFIP/IEEE International
  Symposium on Integrated Network Management (IM)}}. IEEE,
  \bibinfo{pages}{872--877}.
\newblock


\bibitem[\protect\citeauthoryear{Holzapfel, Karl, Lotz, Carle, Djeffal, Fruck,
  Haack, Heckmann, Kindt, K{\"o}ppl, et~al\mbox{.}}{Holzapfel
  et~al\mbox{.}}{2020}]%
        {holzapfel2020digital}
\bibfield{author}{\bibinfo{person}{Kilian Holzapfel}, \bibinfo{person}{Martina
  Karl}, \bibinfo{person}{Linus Lotz}, \bibinfo{person}{Georg Carle},
  \bibinfo{person}{Christian Djeffal}, \bibinfo{person}{Christian Fruck},
  \bibinfo{person}{Christian Haack}, \bibinfo{person}{Dirk Heckmann},
  \bibinfo{person}{Philipp~H Kindt}, \bibinfo{person}{Michael K{\"o}ppl},
  {et~al\mbox{.}}} \bibinfo{year}{2020}\natexlab{}.
\newblock \showarticletitle{Digital Contact Tracing Service: An improved
  decentralized design for privacy and effectiveness}.
\newblock \bibinfo{journal}{\emph{arXiv preprint arXiv:2006.16960}}
  (\bibinfo{year}{2020}).
\newblock


\bibitem[\protect\citeauthoryear{Joshi, Hong, and Katti}{Joshi
  et~al\mbox{.}}{2013}]%
        {joshi2013pinpoint}
\bibfield{author}{\bibinfo{person}{Kiran Joshi}, \bibinfo{person}{Steven Hong},
  {and} \bibinfo{person}{Sachin Katti}.} \bibinfo{year}{2013}\natexlab{}.
\newblock \showarticletitle{PinPoint: Localizing Interfering Radios}. In
  \bibinfo{booktitle}{\emph{10th {USENIX} Symposium on Networked Systems Design
  and Implementation ({NSDI} 13)}}. \bibinfo{publisher}{{USENIX} Association},
  \bibinfo{address}{Lombard, IL}, \bibinfo{pages}{241--253}.
\newblock
\showISBNx{978-1-931971-00-3}
\urldef\tempurl%
\url{https://www.usenix.org/conference/nsdi13/technical-sessions/presentation/joshi}
\showURL{%
\tempurl}


\bibitem[\protect\citeauthoryear{Legendre, Humbert, Mermoud, and
  Lenders}{Legendre et~al\mbox{.}}{2020}]%
        {legendre2020contact}
\bibfield{author}{\bibinfo{person}{Franck Legendre}, \bibinfo{person}{Mathias
  Humbert}, \bibinfo{person}{Alain Mermoud}, {and} \bibinfo{person}{Vincent
  Lenders}.} \bibinfo{year}{2020}\natexlab{}.
\newblock \showarticletitle{Contact tracing: An overview of technologies and
  cyber risks}.
\newblock \bibinfo{journal}{\emph{arXiv preprint arXiv:2007.02806}}
  (\bibinfo{year}{2020}).
\newblock


\bibitem[\protect\citeauthoryear{Li, Hu, Zhong, Tsui, and Chan}{Li
  et~al\mbox{.}}{2021}]%
        {li2021vcontact}
\bibfield{author}{\bibinfo{person}{Guanyao Li}, \bibinfo{person}{Siyan Hu},
  \bibinfo{person}{Shuhan Zhong}, \bibinfo{person}{Wai~Lun Tsui}, {and}
  \bibinfo{person}{S-H~Gary Chan}.} \bibinfo{year}{2021}\natexlab{}.
\newblock \showarticletitle{VContact: Private WiFi-based IoT contact tracing
  with virus lifespan}.
\newblock \bibinfo{journal}{\emph{IEEE Internet of Things Journal}}
  (\bibinfo{year}{2021}).
\newblock


\bibitem[\protect\citeauthoryear{Liu, Yin, Jiang, and He}{Liu
  et~al\mbox{.}}{2021}]%
        {liu2021wibeacon}
\bibfield{author}{\bibinfo{person}{Ruofeng Liu}, \bibinfo{person}{Zhimeng Yin},
  \bibinfo{person}{Wenchao Jiang}, {and} \bibinfo{person}{Tian He}.}
  \bibinfo{year}{2021}\natexlab{}.
\newblock \showarticletitle{WiBeacon: expanding BLE location-based services via
  wifi}. In \bibinfo{booktitle}{\emph{Proceedings of the 27th Annual
  International Conference on Mobile Computing and Networking}}.
  \bibinfo{pages}{83--96}.
\newblock


\bibitem[\protect\citeauthoryear{Malloy, Barford, Alp, Koller, and
  Jewell}{Malloy et~al\mbox{.}}{2017}]%
        {malloy2017internet}
\bibfield{author}{\bibinfo{person}{Matthew Malloy}, \bibinfo{person}{Paul
  Barford}, \bibinfo{person}{Enis~Ceyhun Alp}, \bibinfo{person}{Jonathan
  Koller}, {and} \bibinfo{person}{Adria Jewell}.}
  \bibinfo{year}{2017}\natexlab{}.
\newblock \showarticletitle{Internet device graphs}. In
  \bibinfo{booktitle}{\emph{Proceedings of the 23rd ACM SIGKDD International
  Conference on Knowledge Discovery and Data Mining}}.
  \bibinfo{pages}{1913--1921}.
\newblock


\bibitem[\protect\citeauthoryear{Malloy, Cahn, and Koller}{Malloy
  et~al\mbox{.}}{2021}]%
        {malloy2020digital}
\bibfield{author}{\bibinfo{person}{Matthew Malloy}, \bibinfo{person}{Aaron
  Cahn}, {and} \bibinfo{person}{Jon Koller}.} \bibinfo{year}{2021}\natexlab{}.
\newblock \showarticletitle{Digital Contact Tracing Using IP Colocation}. In
  \bibinfo{booktitle}{\emph{2021 IEEE International Conference on Internet of
  Things and Intelligence Systems (IoTaIS)}}. IEEE, \bibinfo{pages}{73--78}.
\newblock


\bibitem[\protect\citeauthoryear{Manavi, Nekkanti, Choudhary, and
  Jayapandian}{Manavi et~al\mbox{.}}{2020}]%
        {manavi2020review}
\bibfield{author}{\bibinfo{person}{S~Yoshita Manavi}, \bibinfo{person}{Vinuthna
  Nekkanti}, \bibinfo{person}{Ram~Shankar Choudhary}, {and} \bibinfo{person}{N
  Jayapandian}.} \bibinfo{year}{2020}\natexlab{}.
\newblock \showarticletitle{Review on emerging Internet of Things technologies
  to fight the COVID-19}. In \bibinfo{booktitle}{\emph{2020 Fifth International
  Conference on Research in Computational Intelligence and Communication
  Networks (ICRCICN)}}. IEEE, \bibinfo{pages}{202--208}.
\newblock


\bibitem[\protect\citeauthoryear{Mcheick, Hassan, and Kteich}{Mcheick
  et~al\mbox{.}}{2021}]%
        {mcheick2021d2d}
\bibfield{author}{\bibinfo{person}{Hamid Mcheick},
  \bibinfo{person}{Houssam~Hajj Hassan}, {and} \bibinfo{person}{Hanane
  Kteich}.} \bibinfo{year}{2021}\natexlab{}.
\newblock \showarticletitle{D2D Communication: COVID-19 Contact Tracing
  Application Using Wi-Fi Direct}. In \bibinfo{booktitle}{\emph{2021
  International Conference on Smart Applications, Communications and Networking
  (SmartNets)}}. IEEE, \bibinfo{pages}{1--6}.
\newblock


\bibitem[\protect\citeauthoryear{Monroe, Tazi, and Das}{Monroe
  et~al\mbox{.}}{2021}]%
        {monroe2021location}
\bibfield{author}{\bibinfo{person}{Callie Monroe}, \bibinfo{person}{Faiza
  Tazi}, {and} \bibinfo{person}{Sanchari Das}.}
  \bibinfo{year}{2021}\natexlab{}.
\newblock \showarticletitle{Location Data and COVID-19 Contact Tracing: How
  Data Privacy Regulations and Cell Service Providers Work In Tandem}. In
  \bibinfo{booktitle}{\emph{Proceedings of the Workshop on Usable Security and
  Privacy (USEC)}}.
\newblock


\bibitem[\protect\citeauthoryear{Munzert, Selb, Gohdes, Stoetzer, and
  Lowe}{Munzert et~al\mbox{.}}{2021}]%
        {munzert2021tracking}
\bibfield{author}{\bibinfo{person}{Simon Munzert}, \bibinfo{person}{Peter
  Selb}, \bibinfo{person}{Anita Gohdes}, \bibinfo{person}{Lukas~F Stoetzer},
  {and} \bibinfo{person}{Will Lowe}.} \bibinfo{year}{2021}\natexlab{}.
\newblock \showarticletitle{Tracking and promoting the usage of a COVID-19
  contact tracing app}.
\newblock \bibinfo{journal}{\emph{Nature Human Behaviour}}
  (\bibinfo{year}{2021}), \bibinfo{pages}{1--9}.
\newblock


\bibitem[\protect\citeauthoryear{Nguyen, Luo, and Watkins}{Nguyen
  et~al\mbox{.}}{2020}]%
        {nguyen2020epidemic}
\bibfield{author}{\bibinfo{person}{Khuong~An Nguyen}, \bibinfo{person}{Zhiyuan
  Luo}, {and} \bibinfo{person}{Chris Watkins}.}
  \bibinfo{year}{2020}\natexlab{}.
\newblock \showarticletitle{Epidemic contact tracing with smartphone sensors}.
\newblock \bibinfo{journal}{\emph{Journal of Location Based Services}}
  \bibinfo{volume}{14}, \bibinfo{number}{2} (\bibinfo{year}{2020}),
  \bibinfo{pages}{92--128}.
\newblock


\bibitem[\protect\citeauthoryear{Ocheja, Cao, Ding, and Yoshikawa}{Ocheja
  et~al\mbox{.}}{2020}]%
        {ocheja2020quantifying}
\bibfield{author}{\bibinfo{person}{Patrick Ocheja}, \bibinfo{person}{Yang Cao},
  \bibinfo{person}{Shiyao Ding}, {and} \bibinfo{person}{Masatoshi Yoshikawa}.}
  \bibinfo{year}{2020}\natexlab{}.
\newblock \showarticletitle{Quantifying the Privacy-Utility Trade-offs in
  COVID-19 Contact Tracing Apps}.
\newblock \bibinfo{journal}{\emph{arXiv preprint arXiv:2012.13061}}
  (\bibinfo{year}{2020}).
\newblock


\bibitem[\protect\citeauthoryear{Oikonomidis, Fouskas, and
  Vlachopoulou}{Oikonomidis et~al\mbox{.}}{2021}]%
        {oikonomidis2021role}
\bibfield{author}{\bibinfo{person}{Theodoros Oikonomidis},
  \bibinfo{person}{Konstantinos Fouskas}, {and} \bibinfo{person}{Maro
  Vlachopoulou}.} \bibinfo{year}{2021}\natexlab{}.
\newblock \showarticletitle{The Role of Wi-Fi Positioning Systems in Safety
  Against COVID-19}. In \bibinfo{booktitle}{\emph{Strategic Innovative
  Marketing and Tourism in the COVID-19 Era: 9th ICSIMAT Conference 2020}}.
  Springer, \bibinfo{pages}{111--119}.
\newblock


\bibitem[\protect\citeauthoryear{Petrovi{\'c} and Koci{\'c}}{Petrovi{\'c} and
  Koci{\'c}}{2021}]%
        {petrovic2021iot}
\bibfield{author}{\bibinfo{person}{N Petrovi{\'c}} {and} \bibinfo{person}{{\DH}
  Koci{\'c}}.} \bibinfo{year}{2021}\natexlab{}.
\newblock \showarticletitle{IoT for COVID-19 Indoor Spread Prevention: Cough
  Detection, Air Quality Control and Contact Tracing}. In
  \bibinfo{booktitle}{\emph{2021 IEEE 32nd International Conference on
  Microelectronics (MIEL)}}. IEEE, \bibinfo{pages}{297--300}.
\newblock


\bibitem[\protect\citeauthoryear{Roy, Kumbhar, Dhillon, Saxena, Shin, and
  Singh}{Roy et~al\mbox{.}}{2020}]%
        {roy2020efficient}
\bibfield{author}{\bibinfo{person}{Abhishek Roy}, \bibinfo{person}{Farooque~H
  Kumbhar}, \bibinfo{person}{Harpreet~S Dhillon}, \bibinfo{person}{Navrati
  Saxena}, \bibinfo{person}{Soo~Young Shin}, {and} \bibinfo{person}{Sukhdeep
  Singh}.} \bibinfo{year}{2020}\natexlab{}.
\newblock \showarticletitle{Efficient monitoring and contact tracing for
  COVID-19: A smart IoT-based framework}.
\newblock \bibinfo{journal}{\emph{IEEE Internet of Things Magazine}}
  \bibinfo{volume}{3}, \bibinfo{number}{3} (\bibinfo{year}{2020}),
  \bibinfo{pages}{17--23}.
\newblock


\bibitem[\protect\citeauthoryear{Sahraoui, De~Lucia, Vegni, Kerrache, Amadeo,
  and Korichi}{Sahraoui et~al\mbox{.}}{2021}]%
        {sahraouitraceme}
\bibfield{author}{\bibinfo{person}{Yesin Sahraoui}, \bibinfo{person}{Ludovica
  De~Lucia}, \bibinfo{person}{Anna~Maria Vegni},
  \bibinfo{person}{Chaker~Abdelaziz Kerrache}, \bibinfo{person}{Marica Amadeo},
  {and} \bibinfo{person}{Ahmed Korichi}.} \bibinfo{year}{2021}\natexlab{}.
\newblock \showarticletitle{TraceMe: Real-Time Contact Tracing and Early
  Prevention of COVID-19 based on Online Social Networks}.
\newblock \bibinfo{journal}{\emph{Academia.edu}} (\bibinfo{year}{2021}).
\newblock


\bibitem[\protect\citeauthoryear{Sanderson}{Sanderson}{2021}]%
        {sanderson2021balancing}
\bibfield{author}{\bibinfo{person}{Pollyanna Sanderson}.}
  \bibinfo{year}{2021}\natexlab{}.
\newblock \showarticletitle{Balancing public health and civil liberties:
  Privacy aspects of contact-tracing technologies}.
\newblock \bibinfo{journal}{\emph{IEEE Security \& Privacy}}
  \bibinfo{volume}{19}, \bibinfo{number}{4} (\bibinfo{year}{2021}),
  \bibinfo{pages}{65--69}.
\newblock


\bibitem[\protect\citeauthoryear{Saw, Tan, Liu, and Liu}{Saw
  et~al\mbox{.}}{2021}]%
        {saw2021predicting}
\bibfield{author}{\bibinfo{person}{Young~Ern Saw}, \bibinfo{person}{Edina
  Yi-Qin Tan}, \bibinfo{person}{Jessica~Shijia Liu}, {and}
  \bibinfo{person}{Jean~CJ Liu}.} \bibinfo{year}{2021}\natexlab{}.
\newblock \showarticletitle{Predicting Public Uptake of Digital Contact Tracing
  During the COVID-19 Pandemic: Results From a Nationwide Survey in Singapore}.
\newblock \bibinfo{journal}{\emph{Journal of medical Internet research}}
  \bibinfo{volume}{23}, \bibinfo{number}{2} (\bibinfo{year}{2021}),
  \bibinfo{pages}{e24730}.
\newblock


\bibitem[\protect\citeauthoryear{Sun, Liu, Du, Xu, and Wu}{Sun
  et~al\mbox{.}}{2021}]%
        {sun2021mitigating}
\bibfield{author}{\bibinfo{person}{Hao-Chen Sun}, \bibinfo{person}{Xiao-Fan
  Liu}, \bibinfo{person}{Zhan-Wei Du}, \bibinfo{person}{Xiao-Ke Xu}, {and}
  \bibinfo{person}{Ye Wu}.} \bibinfo{year}{2021}\natexlab{}.
\newblock \showarticletitle{Mitigating COVID-19 Transmission in Schools With
  Digital Contact Tracing}.
\newblock \bibinfo{journal}{\emph{IEEE Transactions on Computational Social
  Systems}} (\bibinfo{year}{2021}).
\newblock


\bibitem[\protect\citeauthoryear{Tahiliani, Hassija, Chamola, Kanhere, Guizani,
  et~al\mbox{.}}{Tahiliani et~al\mbox{.}}{2021}]%
        {tahiliani2021privacy}
\bibfield{author}{\bibinfo{person}{Aman Tahiliani}, \bibinfo{person}{Vikas
  Hassija}, \bibinfo{person}{Vinay Chamola}, \bibinfo{person}{Salil~S Kanhere},
  \bibinfo{person}{Mohsen Guizani}, {et~al\mbox{.}}}
  \bibinfo{year}{2021}\natexlab{}.
\newblock \showarticletitle{Privacy-Preserving and Incentivized Contact Tracing
  for COVID-19 Using Blockchain}.
\newblock \bibinfo{journal}{\emph{IEEE Internet of Things Magazine}}
  \bibinfo{volume}{4}, \bibinfo{number}{3} (\bibinfo{year}{2021}),
  \bibinfo{pages}{72--79}.
\newblock


\bibitem[\protect\citeauthoryear{Tang}{Tang}{2020}]%
        {tang2020privacy}
\bibfield{author}{\bibinfo{person}{Qiang Tang}.}
  \bibinfo{year}{2020}\natexlab{}.
\newblock \showarticletitle{Privacy-preserving contact tracing: current
  solutions and open questions}.
\newblock \bibinfo{journal}{\emph{arXiv preprint arXiv:2004.06818}}
  (\bibinfo{year}{2020}).
\newblock


\bibitem[\protect\citeauthoryear{Tang}{Tang}{2021}]%
        {tang2021another}
\bibfield{author}{\bibinfo{person}{Qiang Tang}.}
  \bibinfo{year}{2021}\natexlab{}.
\newblock \showarticletitle{Another look at privacy-preserving automated
  contact tracing}.
\newblock \bibinfo{journal}{\emph{ACM Transactions on Spatial Algorithms and
  Systems (TSAS)}} \bibinfo{volume}{8}, \bibinfo{number}{2}
  (\bibinfo{year}{2021}), \bibinfo{pages}{1--27}.
\newblock


\bibitem[\protect\citeauthoryear{Thakare, Dohare, and Akhtar}{Thakare
  et~al\mbox{.}}{2021}]%
        {thakare2021p}
\bibfield{author}{\bibinfo{person}{Laxmi Thakare}, \bibinfo{person}{Ruchi
  Dohare}, {and} \bibinfo{person}{Nadeem Akhtar}.}
  \bibinfo{year}{2021}\natexlab{}.
\newblock \showarticletitle{P-Tracer: Proximity Detection for Contact Tracing}.
  In \bibinfo{booktitle}{\emph{2021 International Conference on COMmunication
  Systems \& NETworkS (COMSNETS)}}. IEEE, \bibinfo{pages}{116--118}.
\newblock


\bibitem[\protect\citeauthoryear{Trivedi, Gummeson, and Shenoy}{Trivedi
  et~al\mbox{.}}{2020}]%
        {trivedi2020empirical}
\bibfield{author}{\bibinfo{person}{Amee Trivedi}, \bibinfo{person}{Jeremy
  Gummeson}, {and} \bibinfo{person}{Prashant Shenoy}.}
  \bibinfo{year}{2020}\natexlab{}.
\newblock \bibinfo{title}{Empirical Characterization of Mobility of
  Multi-Device Internet Users}.
\newblock
\newblock
\showeprint[arxiv]{2003.08512}~[cs.CY]


\bibitem[\protect\citeauthoryear{Trivedi, Zakaria, Balan, Becker, Corey, and
  Shenoy}{Trivedi et~al\mbox{.}}{2021}]%
        {trivedi2021wifitrace}
\bibfield{author}{\bibinfo{person}{Amee Trivedi}, \bibinfo{person}{Camellia
  Zakaria}, \bibinfo{person}{Rajesh Balan}, \bibinfo{person}{Ann Becker},
  \bibinfo{person}{George Corey}, {and} \bibinfo{person}{Prashant Shenoy}.}
  \bibinfo{year}{2021}\natexlab{}.
\newblock \showarticletitle{WiFiTrace: Network-based Contact Tracing for
  Infectious Diseases Using Passive WiFi Sensing}.
\newblock \bibinfo{journal}{\emph{Proceedings of the ACM on Interactive,
  Mobile, Wearable and Ubiquitous Technologies}} \bibinfo{volume}{5},
  \bibinfo{number}{1} (\bibinfo{year}{2021}), \bibinfo{pages}{1--26}.
\newblock


\bibitem[\protect\citeauthoryear{Tu, Li, Wang, Wanga, and Yuan}{Tu
  et~al\mbox{.}}{2021}]%
        {tu2021epidemic}
\bibfield{author}{\bibinfo{person}{Pengjia Tu}, \bibinfo{person}{Junhuai Li},
  \bibinfo{person}{Huaijun Wang}, \bibinfo{person}{Kan Wanga}, {and}
  \bibinfo{person}{Yuan Yuan}.} \bibinfo{year}{2021}\natexlab{}.
\newblock \showarticletitle{Epidemic Contact Tracing with Campus WiFi Network
  and Smartphone-based Pedestrian Dead Reckoning}.
\newblock \bibinfo{journal}{\emph{IEEE Sensors Journal}}
  (\bibinfo{year}{2021}).
\newblock


\bibitem[\protect\citeauthoryear{Yi, Xie, and Jamieson}{Yi
  et~al\mbox{.}}{2021}]%
        {yi2021cellular}
\bibfield{author}{\bibinfo{person}{Fan Yi}, \bibinfo{person}{Yaxiong Xie},
  {and} \bibinfo{person}{Kyle Jamieson}.} \bibinfo{year}{2021}\natexlab{}.
\newblock \showarticletitle{Cellular-assisted COVID-19 contact tracing}. In
  \bibinfo{booktitle}{\emph{Proceedings of the 2nd Workshop on Deep Learning
  for Wellbeing Applications Leveraging Mobile Devices and Edge Computing}}.
  \bibinfo{pages}{1--6}.
\newblock


\bibitem[\protect\citeauthoryear{Yoo, Kim, Shin, and J~Clayton}{Yoo
  et~al\mbox{.}}{2020}]%
        {yoo2020bim}
\bibfield{author}{\bibinfo{person}{Wonjae Yoo}, \bibinfo{person}{Hyoungsub
  Kim}, \bibinfo{person}{Minjae Shin}, {and} \bibinfo{person}{Mark J~Clayton}.}
  \bibinfo{year}{2020}\natexlab{}.
\newblock \showarticletitle{BIM-Based Automatic Contact Tracing System Using
  Wi-Fi}.
\newblock  (\bibinfo{year}{2020}).
\newblock


\bibitem[\protect\citeauthoryear{Zagatti, Wu, Ng, and Bressan}{Zagatti
  et~al\mbox{.}}{2021}]%
        {zagatti2021large}
\bibfield{author}{\bibinfo{person}{Guilherme~Augusto Zagatti},
  \bibinfo{person}{Tingfeng Wu}, \bibinfo{person}{See-Kiong Ng}, {and}
  \bibinfo{person}{St{\'e}phane Bressan}.} \bibinfo{year}{2021}\natexlab{}.
\newblock \showarticletitle{A Large-scale Disease Outbreak Analytics System
  based on Wi-Fi Session Logs}. In \bibinfo{booktitle}{\emph{2021 22nd IEEE
  International Conference on Mobile Data Management (MDM)}}. IEEE,
  \bibinfo{pages}{236--239}.
\newblock


\bibitem[\protect\citeauthoryear{Zakaria, Trivedi, Cecchet, Chee, Shenoy, and
  Balan}{Zakaria et~al\mbox{.}}{2020}]%
        {zakaria2020analyzing}
\bibfield{author}{\bibinfo{person}{Camellia Zakaria}, \bibinfo{person}{Amee
  Trivedi}, \bibinfo{person}{Emmanuel Cecchet}, \bibinfo{person}{Michael Chee},
  \bibinfo{person}{Prashant Shenoy}, {and} \bibinfo{person}{Rajesh Balan}.}
  \bibinfo{year}{2020}\natexlab{}.
\newblock \showarticletitle{Analyzing the impact of covid-19 control policies
  on campus occupancy and mobility via passive wifi sensing}.
\newblock \bibinfo{journal}{\emph{arXiv preprint arXiv:2005.12050}}
  (\bibinfo{year}{2020}).
\newblock


\bibitem[\protect\citeauthoryear{Zang and Sweeney}{Zang and Sweeney}{2021}]%
        {zang2021building}
\bibfield{author}{\bibinfo{person}{Jinyan Zang} {and} \bibinfo{person}{Latanya
  Sweeney}.} \bibinfo{year}{2021}\natexlab{}.
\newblock \showarticletitle{Building A Collocation Detection System Using A
  Wi-Fi Sensor Array for COVID-19 Contact Tracing in A University Setting}.
\newblock \bibinfo{journal}{\emph{Case Studies in Public Interest Technology}}
  (\bibinfo{year}{2021}), \bibinfo{pages}{168}.
\newblock


\bibitem[\protect\citeauthoryear{Zhang, Pan, Zhang, Champion, Shen, Xuan, Lin,
  and Shroff}{Zhang et~al\mbox{.}}{2021}]%
        {zhang2021wlan}
\bibfield{author}{\bibinfo{person}{Cheng Zhang}, \bibinfo{person}{Yunze Pan},
  \bibinfo{person}{Yunqi Zhang}, \bibinfo{person}{Adam~C Champion},
  \bibinfo{person}{Zhaohui Shen}, \bibinfo{person}{Dong Xuan},
  \bibinfo{person}{Zhiqiang Lin}, {and} \bibinfo{person}{Ness~B Shroff}.}
  \bibinfo{year}{2021}\natexlab{}.
\newblock \showarticletitle{WLAN-Log-Based Superspreader Detection in the
  COVID-19 Pandemic}.
\newblock \bibinfo{journal}{\emph{High-Confidence Computing}}
  (\bibinfo{year}{2021}), \bibinfo{pages}{100005}.
\newblock


\bibitem[\protect\citeauthoryear{Zhao, Wen, Lin, Xuan, and Shroff}{Zhao
  et~al\mbox{.}}{2020}]%
        {zhao2020accuracy}
\bibfield{author}{\bibinfo{person}{Qingchuan Zhao}, \bibinfo{person}{Haohuang
  Wen}, \bibinfo{person}{Zhiqiang Lin}, \bibinfo{person}{Dong Xuan}, {and}
  \bibinfo{person}{Ness Shroff}.} \bibinfo{year}{2020}\natexlab{}.
\newblock \showarticletitle{On the accuracy of measured proximity of
  Bluetooth-based contact tracing apps}. In
  \bibinfo{booktitle}{\emph{International Conference on Security and Privacy in
  Communication Systems}}. Springer, \bibinfo{pages}{49--60}.
\newblock


\end{thebibliography}
\bibliographystyle{ACM-Reference-Format}

\end{document}